# Mitochondrial oxidative phosphorylation: *Debunking the concepts of electron transport chain, proton pumps, chemiosmosis and rotary ATP synthesis*

*Kelath Murali Manoj**

*Satyamjayatu: The Science & Ethics Foundation
Kulappully, Shoranur-2 (PO), Palakkad District, Kerala, India-679122.
Email: satyamjayatu@gmail.com; satyamjayatu@yahoo.com

**Abstract:** The prevailing half-a-century old explanation for cellular respiration solicits a synchronization of electron transfer chain (ETC) and proton pumping activity across the inner mitochondrial membrane, for generating a trans-membrane proton-electro-chemical potential. This outcome is supposedly harnessed by Complex V to synthesize ATP, working via a rotary mode. In the first part of my work (the current write-up), I highlight that a mitochondrion is a highly "proton-limited" microcosm, housing only $\sim10^1$ protons in physiological ranges. It is inconceivable that $\sim10^4$ protein (super)complexes (that exist per mitochondrion) have to deal with protons. A survey of the purported electron transport chain shows that even if we discount the insights revealed by crystal structures and the mandate imposed by Ockham's razor/ evolutionary pressures, more than half of the 54 electron transfer steps (needed to form just one molecule of water!) would be thermodynamically disadvantaged. *In situ*, such a scheme would have little significant kinetic viability. Also, we have a catch 22 in considering that even if the feasible fast intra-protein electron transfers are somehow slowed (in situ, to couple electron transfer with trans-membrane proton translocation), the overall observed water formation rates (achieved at sub-millisecond timescales) cannot be explained with the scheme, owing to the sequential nature of the "electron transport chain". Thereby, it is improbable that "ETC-proton pumps" are operative. Further, it is demonstrated that the chemiosmosis hypothesis was an "accounting logic fail" which sent the scientific community on wild goose chases. Furthermore, based on available facts, it is demonstrated that a rotary synthesis of ATP by Complex V is kinetically and energetically improbable. *In toto*, key incontrovertible facts and rational perspectives are aligned to conclude that the prevailing impressions on oxidative phosphorylation had only served the *fait accompli* in Peter Mitchell's "hypnotic" chemiosmosis hypothesis. In the second part of this work (concomitantly communicated), I have elaborated upon an alternative explanation / mechanism for oxidative phosphorylation.

In aerobic eukaryotes, the final steps of cellular respiration transpire within the mitochondria and the technical term for this process is oxidative phosphorylation (OxPhos). The current perspective on mitochondrial OxPhos (mOxPhos) seeks the orchestration of the following events at/around the mitochondrial membranes (or plasma membrane, as is the case for aerobic prokaryotes):

(i)      An overall downhill (and at times, a roller-coaster type) transfer of electron pairs (except the steps that involve Cytochrome *c*) is relayed from reduced substrates all the way to the electronegative oxygen molecule through a charted "chain/circuit" of large/small molecules ("electron transport chain" or **ETC**). The vitally deterministic sequential scheme (NADH / FADH$_2$ → Complex I / Complex II → Coenzyme Q → Complex III → Cytochrome *c* → Complex IV → Oxygen) comprises of four distinct immobilized multi-protein Complexes I through IV, embedded on the inner mitochondrial membrane and two diffusible species in Coenzyme Q (CoQ) and Cytochrome *c* (Cyt. *c*).

(ii)     The ETC above is efficiently used and coupled with an uphill trans-membrane (matrix to outward) **proton pumping** process, supposed to occur through three Complexes (I, III & IV). Some protons are also taken up into the membrane for CoQ to recycle.

(iii)    The conformation-gated processes above set up a proton-electrochemical gradient across the inner mitochondrial membrane, thereby providing a **chemiosmotic** drive.

(iv)    A subsequent (or synchronous) matrix-ward movement of protons occurs (by virtue of the chemiosmotic gradient set up across the membrane), which is coupled with a **rotary synthesis** of ATP, afforded by a "cyclic-motor" functionality of Complex V or F$_o$ F$_1$ H$^+$-ATPsynthase. This outcome is also supposedly made possible (once again!) by several sophisticated proteins' motional and conformational variations.

The **E**TC-**P**roton Pump-**C**hemiosmosis-**R**otary Synthesis hypothesis (which shall be henceforth called EPCR) seeks an elaborate spatial arrangement of high-fidel electronic transfer circuitry (invoking a spatio-temporal separation of protons and electrons over relatively large dimensions across distinct macroscopic phases), proton pump machineries and molecular motor assemblies. The whole gamut hinges on interlinking of the first two interconnected ideas with the fourth idea by the "chemiosmosis" concept (the crux of the energy transduction logic). This chemiosmosis principle, "the formation of a trans-membrane proton-electro-chemical gradient" was proposed in 1961 by Peter Mitchell [Mitchell, 1961].

Before getting to my story, history first! Peter Mitchell proposed and ardently followed through his now-famous ideas, originally floated as an interesting and alternative conjecture (sans adequate experimental evidence) in 1961. From the 1950s through mid 1970s, the primary hypothesis regarding mOxPhos professed a chemical coupling step sponsored by high energy enzyme intermediate [Slater, 1953]; and the requisite was almost (erroneously!) served for the community [Boyer, 1963]. However, soon enough, key experimental evidence had supposedly accrued that vouched for a proton-pump mechanism [Jagendorf & Uribe, 1966; Reid et al, 1966; Liberman et al, 1969]. Finally, an experimental chimaera using bacteriorhodopsin [Racker & Stoeckenius, 1974] had apparently sealed the debate in Mitchell's favour, leading to his novel ideas being recognized with a Nobel. Thus, the chemiosmosis hypothesis attained a global stamp of approval and very quickly, it acquired a "theory" stature for some researchers [Brand & Lehninger, 1977]. Slater had vehemently disagreed with Mitchell first [Slater, 1967] and then made a volte-face; falling prey hook, line and sinker to Mitchell in the later 1970s [Boyer et al., 1977]. He had recanted upon this revised stance a decade later, well post Mitchell's Nobel recognition and ideological reconciliation [Slater, 1987], and urged researchers on to rethink beyond the available explanations for mOxPhos. By this time, Slater was quite aware that his own idea of enzyme-bound high-energy intermediate had been rendered redundant because Paul Boyer had also reverted by then and floated his rotary synthesis paradigm [Gresser et al, 1979]. Slater was a meticulous and respectable man of science [Borst, 2016]. It is highly unlikely that he would venture out on a vain egotistical path, opening up a cold sore, for settling some old scores. He passed away last year [Borst, 2016], without receiving a befitting response to his call. Not just Slater, other seasoned researchers have also doubted the EPCR explanation and called for a more chemically viable mechanistic view [Wainio, 1985].

Ever since I graduated in biology (1990), I could not find answers to some elementary doubts I had on mOxPhos [Please refer to Item 1, Supplementary Information for my concerns then.] and therefore, I continued to see EPCR as an incongruous bunch of ideas. In my postdoctoral research, over the past two decades, I made some discoveries in the domain of heme/flavo enzymes and they could be explained only by challenging long-standing impressions on redox enzyme catalytic mechanisms and electron transfer phenomena [Manoj et al, 2016a-d; Venkatachalam et al, 2016; Parashar et al, 2014a-b; Gade et al, 2012; Gideon et al, 2012;

Parashar & Manoj, 2012; Manoj et al, 2010a-b; Manoj & Hager, 2008; Manoj, 2006]. In recent times, the crystal structures of the last remaining element in the mitochondrial ETC (Complex I, [Zhu et al, 2016; Stroud et al, 2016, Fiedorczuk et al, 2016]) and some supra-complex assemblies' organization [Gu et al, 2016; Wu et al, 2016] were revealed. The insights from all these works cited above challenge some long-held assumptions (for example- the highly circuitous and intricate Coenzyme Q recycling process, which was deemed integral to the ETC mechanism [Mitchell, 1975; Zhang et al, 1998]). Such developments confirmed my doubts and further flamed my curiosities. Since facts are the only authorities in science and everything else can be challenged, I ventured to critically dissect the prevailing explanations on mOxPhos and put forth a facile explanation based on "murburn concept" [Manoj et al, 2016a-d; Venkatachalam et al, 2016], which has more spatio-temporal, chemical and evolutionary appeal. In this manuscript (the first part of my work), I debunk the available EPCR explanation for mOxPhos.

**Results & Discussion**

**1. The proton "trump":** The simplest and strongest argument against the EPCR hypothesis rests in accounting protons. We know that a mitochondrion is a bean shaped (or cylindrical, the simplest geometric approximation, with linear dimensions of 0.5 μm x 1 μm) organelle with a volume of ~0.2 μm$^3$. Now, 1 μm$^3$ equals $10^{-15}$ L and we have $10^{-7}$ M concentration of protons at neutral pH. Therefore, the number of protons in a mitochondrion would equal-

$$= [(6.023 \times 10^{23} \times 10^{-7}) \times (0.2 \times 10^{-15})] \approx 12 \text{ or } (\sim 10^1)!$$

Now consider two facts- (i) most mitochondria would have lower volume terms (because of cristae invaginations), and (ii) ATP synthesis has been noted at higher pH (at much lower proton concentrations). Then, the proton pump concept has absolutely no *locus standi*! Though the distribution density of the respiratory Complexes would be ~ $10^3$ - $10^4$ per mitochondrion [Gupte et al, 1984; Schwerzmann et al, 1986], even a single "ETC battery" won't find enough protons to pump out, at any given instant. I hope that this simple trumping argument (*The fact-mitochondria are highly proton-deficient microcosms!*) would give the reader ample confidence to challenge the remaining "foundations" of EPCR. Actually, this paragraph is enough to dismantle EPCR *in toto*, but since we are dealing with a long-entrenched belief paradigm, we must indulge in component-wise, methodical debunking.

**2. Correlating structural features with the overall functional attributes:** First, I shall compare mitochondrial functioning with a man-made machine (automobile), to highlight the dearth of components and the lack of ordered arrangement (modularity) that the EPCR explanation solicits. Mitochondria must possess the following structural and spatially demarcated modular entities-

(i) bio-engine (ETC: burn the fuel NADH to generate the energy to do the work, retain difference of electron accepting-donating redox couples across a complex, contiguous and compartmentalized "organic" circuitry),

(ii) bio-dynamo (Proton pumps: generate a proton-electro-chemical gradient across the inner membrane),

(iii) bio-battery or bio-power plant (ATPsynthase: serve as the mechano-energetic coupler to cyclically synthesize the energy currency, which rests the ability to do chemical work), &

(iv) bio-sensor & bio-regulator/pacemaker (Chemiosmosis logic: self-analyze and self-regulate proton concentration, movements of electrons, tap into electrical field, etc. and thereafter, govern the movement/synchronization of molecular motor).

Though theoretical treatments of mitochondrial modularity are available [Gowthrop, 2017], there is hardly any experimental or practical evidence of ordered arrangement of the overall components within the mitochondria. (The presence of "supra-complexes / respirasomes" or sophisticated mechanisms of "Rieske protein flipping" don't suffice to achieve the needful either!) While the automobile achieves the functional mandate with distinct demarcation of the different modules (when at a given a starting point, both the battery and the fuel serve as two independent sources of energy), there is no such scope in the mitochondria. The compartmentalization of mitochondria into two parts alone (matrix and inter-membrane space) and the simple facet of "proteins being distributed on/across a bi-lamellar partitioning" affords too little scope for the modularity and their functional synchronizations. The structural outlay and the minimal aspects of the overall reaction scheme involving the 6 proteins/complexes involved in the EPCR scheme of mOxPhos is given in Figure 1. The distribution ratio for Complexes I : II : III : IV : V : Cyt. *c* is 1 : 2 : 2-3 : 6 : 6 : 6-9 [Gupte et al, 1984; Schwerzmann et al, 1986]. It is inexplicable why Complexes I & II should be so low, why Complex IV should be so high and why Complex IV : Cyt. *c* ratio is ~1 : 1-1.5. By all means, the prevailing ETC is highly unlikely to work in a practically sequential/synchronized scheme.

Very importantly, the structural features of the Complexes I through IV [Sazanov, 2015; Yankovskaya et al, 2003; Iwata et al, 1998; Tsukihara et al., 1996] do not show any specific or special features that fits the mandate of EPCR. There is no evidence for any complex circuitry within the complexes and one can find little direct experimental evidence for trans-membrane pumps [Natecz, 1986]. The recent respirasome structure [Wu at al & Gu et al, 2016] shows that Complex I's horizontal arm is a lipid-anchoring domain for the trans-membrane portion of Complex III. If we imagined the huge bulbous protrusions of the Complex III dimer as guarding gates of Complex I's pumps, thereby functioning as a "megapump" through the cores of both the complexes, we know that it would just lead to another "sucking up" exercise that wouldn't pump anything (but it could only pander to the dictates of EPCR!). Calling trans-membrane helices (TMHs) of such Complexes as pumps is unjust, because pumps usually use up ATP. These Complexes' TMHs could perhaps qualify for a transport portal for protons, but it would be much slower than channels/pores or "real pumps". Further, I could never comprehend how ATPsynthase could gauge proton concentration or electro-chemical gradient across the membrane and tap into the same. In man-made motors/generator (cyclic functions employing electricity/magnetism), it is done through electromagnetic induction. I couldn't see or envisage any such equivalents in the mitochondria.

Figure 1 shows that the matrix-ward projection of the five complexes are roughly ordered as follows- Complex IV (~3-4 nm) < Complex II (~8 nm) < Complex III (~8 nm) < Complex I (~14 nm) < Complex V (~15 nm). Only Complexes III and IV have any significant projections into the inter-membrane space. While the outward projections of Complexes III & IV can be explained by the EPCR hypothesis (so as to serve for binding the oxidized and reduced forms of Cyt. *c* respectively), the matrix-ward projections of Complexes I, II & III are quite a puzzle. Complex III- has no known substrate to bind at the matrix side, but yet, it has a big bulbous protrusion into the matrix. Further, why did evolution not tuck away the redox centres of the flavoenzymes (Complexes I & II) into the trans-membrane domain and choose not to send out projections into the matrix? The locations of several redox centres are out of the proposed electron transfer routes and several redox centers are located too far apart for effective ET. Further, the proposed routes of electron transfer are many times situated away from the purported proton pump routes. This is clearly evident in Complex I where the matrix-ward protruding arm is the proposed route for electron transfers and the membrane-embedded trans-membrane helices are the purported proton pumps. (A detailed analysis of these aspects follows in section 4.) In such cases, how is the ET process coupled to proton pumping? The confusion in the field is evident from the fact that I could come across at least half a dozen explanations for the purported proton pumping mechanism of Complex I. (If someone is interested in the redundant concept, they may peruse some of the citations listed in Treberg & Brand, 2011].) One could accept a mechanistic proposal that an electron's movement through a given route in a membrane could concomitantly translocate one proton in any given direction. In the ETC setups, the complexes are supposed to take in and pump several protons. This would necessitate non-overlapping electron circuit loops within the trans-membrane domain and there is little evidence for such unrealistic requirements. Further, some redox centers within the Complexes are located at least more than 10 Å away from the matrix milieu, as seen in Complexes III and IV. The only way the reduction of a redox centre could be associated with a concomitant proton uptake from the matrix side would be if water/proton channels could be present through the protein. If a water channel traversed that far into the protein/membrane and if the proton were to be thrown out into the intermembrane side (from this midpoint), then, for all practical matters, we would have a functional proton conduit within the membrane. This would negate the assumption that the inner membrane is impermeable to protons (and such a membrane would have very low resistance, as opposed to Mitchell's assumptions/requirements).

**3. Reaction stoichiometry:** The exact amounts of protons consumed or pumped out seemed to vary (as one goes through a chronological account) for each one of the Complexes and so did the oxygen consumed with respect to phosphate incorporated into ADP. Both of these moieties (protons & phosphate) are stable and accountable species; their operations and stoichiometries in chemical reactions should give a whole number and constant relations with the other participants within the system, given the perfectly deterministic schema of EPCR. While the researchers have been trying to arrive at a consensus regarding this requisite, their experimental data seemed to speak the other way, whether it is P:O (the ATP molecules formed to oxygen atom consumed, when a given substrate like NADH or succinate was presented) or H:P (the number of protons required for making one molecule of ATP) ratios [Rottenberg & Caplan, 1967; Lemasters, 1984; Hinkle 2005; Brand & Lehninger, 1977; Turina et al, 2003; Steigmiller et al, 2008; Petersen et al, 2012]. I wonder if the convergence to the required numbers (as promulgated by the EPCR hypothesis and as revealed to suit the crystallographic data that came out in due time) that we see in modern times resulted more owing to an "enforced aesthetic outlook", than emerging out of experimental consensus. Also, the protocols used to arrive at these numbers leave a lot desired. The murky data in the field (particularly, in 1950s through 1970s, when the chemiosmosis hypothesis had not gained unquestionable credentials) does not agree with ETC as the primary electron route. Further, trying to compare reaction stoichiometries with protein distributions (discussed in earlier section) does not afford any intelligible correlations either.

**4. Evaluating the feasibility/utility of ETC:** Before we proceed, an update on the foundations of redox biochemistry is warranted. A simple understanding of chemical principles tells us that electron donating and accepting ability within a micro-heterogeneous regime (as the mitochondria is envisaged to be) is dependent primarily upon the key chemico-physical attributes of - **redox potentials, relative concentrations, mobility/distances, partitioning and stability** of the participating/resulting entities. Unlike other reactions where affinity based bindings are obligatory for determining efficient catalysis and outcomes, redox processes (particularly involving one-electron transfers) do not need that a perceived donor-acceptor pair meet and bind within the milieu [Manoj et al, 2016a-d; Gade et al, 2012; Parashar & Manoj, 2012, Manoj et al, 2010a; Manoj, 2006]. For example- Cyt. *c* may get an electron from say A (directly or indirectly) and retain it at a given status within the milieu. If the prevailing stature changed and/or if a more suitable/stable electron acceptor B entered the milieu, Cyt. *c* cannot hold on to the electron anymore (even though it might not have affinities for B).

In the light above, the electron transfer machinery espoused in the current system is too "convoluted" to serve as the primary mode of electron transport/transfer in the system. If the relay of electrons is indeed in the fastidious sequence of (NADH to Complex I to CoQ to Complex III to Cyt. *c* to Complex IV to oxygen) or (Succinate to Complex II to CoQ to Complex III to Cyt. *c* to Complex IV to oxygen), how is the wiring achieved with respect to the common portal of Complex III? [Please refer Item 2 of Supplementary Information for a critical commentary into the purported functioning of CoQ cycle.] Since there are no compartmentalizations, modularity or synchronization machineries, the system must work on the terms of probability alone. Therefore, let's see the "probability" considerations of the overall ETC setup.

***a. The omnipresence of oxygen:*** Under normal/physiological conditions, oxygen is available at $10^1$ to $10^2$ μM concentration in the aqueous cytoplasm and is at least 4-5 folds more soluble in lipids. Why should such a freely diffusible and highly mobile small molecule stay stuck to the reaction centre of Complex IV alone? Why cannot the electrophilic oxygen (whether triplet, or dynamically rendered singlet in the vicinity of the metalloenzymes/DROS) and the DROS by-products meet the electrons/protons elsewhere, in the inter-membrane space or within the membrane? The lipid bilayer's permeability values of protons/hydroxyl ions, water molecule and gaseous oxygen are in the range of ~$10^2$, ~$10^4$ and >$10^7$ nm s$^{-1}$ respectively [Milo & Phillips, 2015]. It is known that the diffusion coefficient of protons/hydroxyl ions, water and oxygen in aqueous systems are in the range of $10^9$ - $10^{10}$ nm$^2$ s$^{-1}$. Therefore, the small dimensions, amphipathic nature and high motility of oxygen-centred entities (as exemplified by singlet or triplet oxygen, superoxide or hydroxyl radicals, hydroxide ion, etc.) could easily afford them a linear dimensional coverage radius of ~$10^1$ to $10^3$ Å/μs in the phospholipid interface, reaching out even into some occluded redox centres of proteins. Since effective electron transfer phenomena observed in the biological systems occur across a distance of ~10 Angstroms in micro- to milli- second time scales, oxygen-based small entities would be "practically everywhere at most time points" (with respect to the spatio-temporal considerations relevant to the reaction realm). There could be no ordering or deterministic scheme that could outdo this probabilistic predicament. There are no factual evidences to show that Complexes I through IV and CoQ/Cyt. *c* are spatially arranged in a specific/sequential manner at the required distribution densities, to afford significant connection or collision frequencies on the mitochondrial membrane. Therefore, no circuitry within/across the phospholipid membrane could circumvent oxygen from shunting the proposed circuitries. Oxygen can receive one, two, three or four electrons (or hydrogen atoms), to respectively form superoxide, peroxide, hydroxyl radical + hydroxide ion and water (two molecules). [Please refer Item 3 of Supplementary Information for an overview of some simple oxygen-based species that could be encountered in the system.] It is known that flavins can spontaneously activate molecular oxygen to give superoxide, Fe centers can produce hydroxyl radicals and Cu is an efficient singlet oxygen generator. Besides, conversion of oxygen centered species (in such metallo-flavin laden systems) from triplet to singlet or doublet statures is also facile. No one will contest that such type of diffusible and reactive oxygen species (DROS) are formed in the relevant redox system (*in vitro*, *in situ* and *in vivo*) and there is a ton of evidence for this fact [Grivennikova & Vinogradov, 2006; Drose, 2013; Bleier & Drose, 2013]. Besides, there exists no real physical contiguity (wiring!) in the ETC because CoQ and Cyt. *c* are freely soluble species within the electronic circuitry, both of which are functionally and spatially demarcated. CoQ is purportedly a two-electron relay agent within the inner phospholipid membrane whereas Cyt. *c* is a one-electron relay agent found solubilized within the inter-membrane space of the mitochondrion. All this means that the so-called ETC circuitry would be easily broken, it cannot prevent DROS formation, and therefore, there would be little aesthetic utility in having such a "circuitry" anyway.

***b***. *The specificity of electron transfer machinery:* If the electron transfer machinery was so specific (as it is currently believed to be), how could man-made synthetic organic molecules/ dyes/ radicals serve as electron donors and acceptors in the ETC? Clearly, affinity binding based rationale is not on the high priority list for the ETC redox reaction logic. Why do anaerobic life forms die if they are exposed to oxygen? The simple answer therein must be that oxygen shunts all electrical systems by serving as the ultimate electron acceptor (at all points within the redox system). Then, why should it not be so in the aerobic system too? (Quite simply, it is!)

***c***. *The commitment of oxygen to Complex IV:* The most vital critical inquiry rests on the premise that it is impossible to envisage the oxygen tethering mechanism putting on four electrons and four protons to the diatomic oxygen at one stroke or via a series of steps. It must be noted that the ETC fails totally if the oxygen's bound presence within complex IV is not a highly fecund and tight binding process. It is well-known that binding of a diatomic gaseous molecule like oxygen (or CO/NO or even an ion like cyanide) to a metal centre is a temporal "on-off" (binding-detachment) process dictated by thermodynamic equilibrium. Usually, the amount of time the ligand is "off" is greater than the amount of time it is "on". (FTIR/Raman Spectroscopy experiments with several proteins' metal centres indicate that presumed tight-binding diatomic ligands get displaced very easily, even at low energy microcosms of -100 °C.) It is inconceivable that a molecular tethering mechanism could evolve or be applicable for all of the different species (the combinations of diatomic oxygen ± one to four electrons ± one to four protons) that could potentially be formed on that reaction centre with the purported mechanism. Then again, if indeed a highly versatile and evolved mechanism for binding did exist, why do very low levels of diverse molecules and ions [a hetero-/di- atomic molecule like CO, which strongly binds to Fe(II), in a manner that could simulate oxygen binding OR multi-/hetero- atomic ion like $N_3^-$ and $CN^-$, which bind to Fe(III)] mess with the respiratory logic? Did the system evolve better for the latter toxic molecules/ions too? Repeatedly, how can the small diatomic molecule of oxygen stay bound to both Fe(II) and Fe(III) centre(s) that could be spontaneously formed in the system, without detaching as incompletely reduced species (such as superoxide, peroxide or hydroxyl radical)? For the high rates of water formation, how are four protons (available at ~10 per mitochondrion) relayed into the phospholipid membrane (where the reaction centre is located)?

*d. Analysis of electron transfer steps and their rates:* Since the pseudo-first order rate of ATP formation/hydrolysis by Complex V approaches ~$10^3$ s$^{-1}$, one can gather that the rate of well-coupled ET (say, for one 4-electron relay across a set of Complex I – CoQ – Complex III – Cyt. *c* – Complex IV) must be $\geq 10^3$ s$^{-1}$. It is known that transfer of single electron across two redox centres (the various combinations like: flavin→FeS, FeS↔FeS , FeS↔CoQ, FeS→heme, CoQ↔heme, heme↔heme, heme↔Cu, etc.) across favorable potentials and variable distances would be a micro- to milli- second phenomena. For example, the electron transfer within Complex I (from the FMN, all the way to the last Fe-S centre, N2) was experimentally found to occur in ~$10^2$ μs range [Verkhovskaya et al, 2008]. (We really cannot ascertain that this did occur through the "wiring", but that is beside the point.) One can clearly see that such an elaborate ETC circuitry cannot achieve the ET rate that is actually observed in functional systems. Why?

The concentration of the circuitry components (Complexes I through V and Cyt. *c*; all added up) are in the range of $10^{11}$ to $10^{12}$ cm$^{-2}$ on the inner mitochondrial membrane. (In terms of volume, let's say that this would translate to roughly 25 - 250 μM of each, which is not a conservative estimate, by any means.) Now, let's take the ETC's multi-molecular sequential reaction scheme involving at least 13 collisions/interactions of 7 participants (NADH-**1**-Complex I-**1**-CoQ-**2**-Complex III-**4**-Cyt. *c*-**4**-Complex IV-**1**-O$_2$), each in the range of 10$^{-4}$ M. Let's not forget that protons are also involved in each one of the 13 steps and they are available only at 10$^{-7}$ M. Let's forget the poor mobility of the bulky species involved and interfacial partitioning issues, and graciously assume a second order diffusion limitation regime of 10$^8$ (for proteins) – 10$^9$ (for the "unavailable" protons) M$^{-1}$ s$^{-1}$. (If we involve two-dimensional calculations, then we would also need to factor in the poor mobilities!) Elementary calculations show that a single step involving 10$^{-4}$ M protein would maximally give a pseudo-first order rate of 10$^4$ s$^{-1}$. A single step involving 10$^{-7}$ M protons would give a pseudo-first order rate of 10$^2$ s$^{-1}$. In each one of the 13 steps, these two processes are involved and therefore, each step would be limited by proton availability. So, from elementary analysis, we can conclude that 13 such sequential steps (each with a frequency of 10$^2$ hertz) cannot work hyper-concertedly to give an overall frequency exceeding 10$^3$ hertz. Particularly, some reactions in the ETC scheme are two-electron or four-electron transfer steps (NADH/Succinate to Complex I/II, Complex I/II to CoQ, CoQ to Complex III, Complex IV to O$_2$), which can be envisaged to be relatively slower than the one-electron process. Further, one full cycle of CoQ (the most crucial diffusible component of the phospholipid component of ETC) at Complex III simultaneously requires 2 fully reduced CoQ, 2 protons from the matrix side and two Cyt. *c* from the inter-membrane side (not to mention, specifically chartered movements of Fe-S protein within the Complex III in a very precisely coordinated time scale). After this step, four separate molecules of Cyt. *c* must deliver electrons to Complex IV and these cannot be an instantaneous process too. Fundamentals of reaction chemistry would tell us that such highly super-coordinated multi-molecular reactions definitely pose very high spatio-temporal dictates. These types of multiple hyper-concerted events cannot occur spontaneously and repeatedly and cannot add up to overall conductions within micro/milli- second time frames, particularly given the concentration ratios of reactants. [To get a quantitative physical idea so as to re-establish the inferences drawn above, please browse the analysis of the individual elements of ETC, as given in Item 4 of Supplementary Information.] Now, the data given in Table 1 would summarize the ETC.

While NADH and succinate could potentially deliver two electrons to the Complexes I and II systems through the intermediacy of the bound flavin cofactors and the adjacent wiring, it remains an enigma as to how CoQ receives and gives two one-electron equivalents FeS Complexes (in Complexes I/II and Complex III respectively). Further, as emphasized earlier, it is a totally unthinkable idea to me that a molecule of oxygen would stay bound to Complex IV, waiting indefinitely for four molecules of Cyt. *c* to sequentially ferry the required number of electrons. As per the prevailing ideas, the reduction of one molecule of oxygen at Complex IV (by a total of four electrons derived from a molecule each of NADH and succinate) minimally solicits the synchronous and tandem working (or continuous linking) of ~70 proteins/small molecules present on/across the phospholipid membrane. This ETC solicits that 24 redox active participants must make >54 electron transfers (in batches of one or two electrons) across a collective path of ~600 Å (the minimal conservative distance that 4 electrons must travel from NADH/succinate to $O_2$) within the protein networks alone. If we start with NADH as the sole reductant and include a minimal distance that CoQ and Cyt. *c* would have to commute within the inner membrane and inter-membrane space respectively (and also factor in the distance for CoQ to recycle), we must accept that each one of the electrons must undertake a journey of >$10^3$ Å across a predominantly/relatively low dielectric zone. As I have already shown that theoretically, this hyper-concerted feat cannot be achieved, the practical magnitude of the event should further consolidate the inference drawn. *This statement becomes more relevant when you take into account the barriers provided by the existence of several unfavourable redox potential gradients and many instances where the two adjacent redox centres are > 12 ± 2 Å away.* By a conservative estimate, more than half of the 54 steps could be deemed unviable with respect to the reaction time scales. I envisage that only if both the cumulative distance and the number of discrete electron transfer steps were lower by an order of magnitude, would there be any chance of spontaneously and repeatedly arriving at the experimentally observed water formation rates.

It is very important to see that though electrons do travel against the expected redox potential gradients in solution chemistry, it is only feasible when the concentrations permit them to. That is- the entity with a lower redox potential can also receive electron from a higher potential species. But for that to happen, the concentration of the latter should be several folds or orders higher than the former, even for a difference of few mV. (Else, we must immobilize them and apply an external potential.) In the ETC, this is highly unlikely to occur under the current scheme because each redox couplets is at 1:1 ratio, within the protein complex. Further, several researchers perceive the redox centers' connectivity within the proteins as a "wire" (as exemplified by the connectivity drawn within the finger projections of Complexes I & II or the relay within Complex III for CoQ interaction). This is a very erroneous perception because a wire "conducts electrons" when/where there is a potential difference. Potentials are created or exist, when electrons accumulate at a given point. If there is no applied field, no electrons would flow through a copper wire and electrons would flow in such a wire in a direction based on the applied field. But this is not supposed to occur in the case of our system, where electron flow must be unidirectional. Please take the case of transfers in steps 5 and 7 of Complex I (Table 1 of Item 4, Supplementary Information, highlighted with asterisk). Data from Bridges et al [2012] gives pretty conclusive support to what I am trying to point out. The redox centres N5 and N6b were found oxidized, even under highly optimized "reducing" conditions. This is because they are flanked by two "unfavorable" steps. A re-reading and re-interpretation of Figure 7 (and its legend) of their article should be sufficient to drive home what I pointed out now. There, one can also gauge how the redox potential is dependent on the dielectric of the ambiance. I must say that assuming an $\varepsilon$ value of 4 (so that the ETC may work!) is a very bad idea. The value of 20 seems reasonable; and under which scenario, as expected, the wire analogy does not work.

The thermodynamic drive for the electronic circuitry in EPCR scheme cannot be a push of electron(s) sequentially from lower to higher redox potential proteins because this supposition is not supported by experimental findings [Manoj et al, 2016b]. The distances are often too large and difference of potential too small (at times, even unfavorable!) for the proteins to efficiently relay electrons across the redox centers (through intra or inter molecular transfers between the series of donors and acceptors). The thermodynamic drive cannot be a pull either, exerted by the oxygen bound at Complex IV, via the diffusible species of Cyt. *c*, through the membrane-bound Complex III, and through the membrane-soluble CoQ, all the way until Complexes I or II (because of a lack of continuity in the circuit, as mentioned earlier). Such an ETC is untenable!

Therefore, the current ETC, for all practical purposes, is untenable as an efficient modality for unidirectional relay of one, two, or four electrons (leading to water formation). The prevailing

ETC concept does not have any explanation for the 3 "non-functional" redox centers seen in Complexes I and II. The ETC was supposed to prevent the formation of ROS but we know that it doesn't. Therefore, the ETC scheme is not just improbable, it is redundant. On one end, the EPCR seeks the macroscopic perspective of physical separation of protons and electrons in space and time. At the other end, it requires the very protons and electrons to be drawn together with great abundance and absolute accuracy/precision at certain defined loci/times alone, all along the way. I find all this to be a highly improbable course of events and one that cannot be repeatedly orchestrated with any fecundity. Therefore, one sees little rationale (other than wishful thinking!) for such a hyper-efficient and functional "ETC" to exist on the mitochondrial membrane. The rates achieved by such an "ETC" cannot explain water formation either. Therefore, it is conclusively inferred that such an elaborate "ETC" cannot be "the preferred/probabilistic route".

**5. Assessing the (quantitative) logic of chemiosmosis:** Now that the ETC and proton pump elements of EPCR stand debunked, let us get to the core "regulatory" element, the quantitative logic of the assumptions and explanations of chemiosmosis. The limitations of modern scientific communication do not permit me to demolish the chemiosmosis hypothesis in a tauntingly circuitous route. But I believe that it is a must in this context and therefore, to strike a midpoint, a write-up comparing chemiosmosis with Escher's imaginary waterfall and a realistic man-made watermill (and some other analogies) is included in Item 5 of Supplementary Information. I urge the reader to peruse the same before moving on to the quantitative analyses below.

*Quantitative aspects of chemiosmosis:* Let us "concede" that Mitchell's chemiosmotic mitochondrion is a practically working machine and model it. We can envisage that several functional couplets of [g1-h1], [g2-h2], [g3-h3] etc. could be spread out through the inner membrane and let us also imagine that they work in perfect synchrony. The couplet would comprise at least one module of g (ETC element + outward proton pump) and a complementary module of h (inward proton conduit incorporating the ATP synthesizing work element). The pictorial representation of this minimal idealized mode is rendered in Figure 2.

Now, disregarding all equilibrium and energetics aspects involved, let us assume that there are enough concentrations of the module couplets and there are also ample spatio-temporal opportunities for the ETC-proton pumping machinery to pump out various extents of the initial matrix proton concentration, within a miniscule time $\Delta t_i$ [where $\Delta t_i = (t2 - t1)$ and $\Delta t_i \rightarrow 0$]. Next, let this process be fully reversible in the second phase wherein the protons return into the matrix in yet another miniscule time $\Delta t_f$ [where $\Delta t_f = (t3 - t2)$ and $\Delta t_f \rightarrow 0$]. It is understood that $t1 < t2 < t3$. Now, let us analyze the temporal variations of trans-membrane potentials under some physiologically relevant scenarios (pH 6 to pH 8). The trans-membrane Nernst potential at any point in time, resulting out of proton gradient (under normal biological conditions) would be given by the simple equation: $E = 61 \log [_oH^+] / [_iH^+]$; and the values are given in Table 2.

We can see that temporal status t2 in the scenario 5 (where the initial out:in ratio of proton is 100) is the most favorable picture that the chemiosmosis hypothesis could aspire for. We can see that a potential difference owing to proton pumping results only if the pumping machinery is highly efficient (in a highly coordinated/synchronous scenario). A simple theoretical simulation tells us that even in such a scenario (wherein we start off with a 100 fold concentration difference between matrix and outside), at least 90% of the internal proton concentration should be pumped out for getting the transient potential reach a value of ~180 mV, the threshold required/noted for viable ATP synthesis [that is- for ATP/ADP ratio to exceed 1; Nicholls, 2004]. Starting with a prior gradient of two pH unit difference is also inadequate. If the operational criterion would be to achieve a pmf difference of 180 mV between the two temporal points t2 and t1 (a sort of alternating temporal potential), then too, 90% pumping out becomes essential (as seen with case 9). Scenario 2 is a more realistic one, and it requires the pumping of ~99% of internal protons, to achieve the threshold voltage. This means that the pumps would have achieved their targets only if there was no proton left within the matrix! See that even if the proton concentrations were reverse, theoretically, one could achieve the potentials that Mitchell wanted. So, if the pumps were good at what they did, how could we justify polarity (unidirectional modality) of the system? (If we grant bi-directionality to ATPase or the h module, how can the unregulated g module pumping machinery know when to stop pumping protons?) Having understood the inconsistency of chemiosmotic logic with the idealized model (in the best case scenario), it is now opportune to de-/re- construct it and bring it to a more realistic plane. Let us revoke the assumption that all proton pumping activities are temporally synchronized. Then, no matter what the scenario, an asynchronous proton-pumping exercise of the various modules (this would be the natural case because there are no known global pace setters within the mitochondria) will afford a transient transmembrane proton potential gradient that would be much lower than the most optimistic ones projected in the table. Therefore, the pumps themselves would be assigned the "sin" of dissipating potentials (by asynchronous working), the very attribute given to uncouplers like dinitrophenol. [[Explaining their effect was supposed to be a major success of Mitchell's idea but you can see that it is just yet another misunderstanding.]]

Please see that if synchronization of the pumps is not important in the system, the system can only tap the "energy" of the initial ΔpH, and cannot work in a continuous and dynamic process. In that case, I hope you see that there is no logic for attributing any importance to a matrix to inter-membrane proton pumping machinery at all! (As explained earlier, modularity is very important to achieve a particular function that EPCR sets to achieve.) Further, it is important to see the manifestation of the very simple energetic logic (as exposed earlier) in the practical consideration that-

1. If the inner membrane is non-permeable, the pumping energy would be independent of the pre-existing proton gradient. In which case, work must be done for both pumping out and pumping in, making the proton pumping a totally futile exercise.

2. If the inner membrane has only electrical connectivity across the two macroscopic phases and is "conveniently directional" (permits only an energy-aided "in to out" pumping of protons and a facile "out to in" return), then the energy expended for "in to out" proton mass transfer would concomitantly increase with any incremental concentration or potential gradient that needs to be set up. Further, the buildup of such a higher potential cannot happen if the "out to in" proton movement is facile and does not incur energy expense (that is- if there is no resistance to the inward movement of protons). As soon as a proton goes out expending energy, the proton would return spontaneously. So, the inward proton movement must also be restrictive. This scenario would be countered with the outer-membrane's permissivity. If the inner membrane becomes restrictive, then the protons escape through the outer membrane, thereby dissipating the potentials. In both scenarios, the pumping-based harnessing of an "accumulated or generated" potential difference would become non-viable.

Now, it is known that in biological systems, the potentials of most resting plasma membranes are usually found to range between -40 to -80 mV (the inside being usually negative), and it is generally understood that these potentials are primarily dictated by potassium or chloride ions' permeability/concentrations. These ions are found (in or out) at concentration ranges of $10^{-3}$ to $10^{-5}$ M. The plasma membrane potential is generally not modulated by proton permeability or concentrations because protons are available in the range of only $\sim 10^{-7}$ M. Now, you can see that if the resting potential is in some small way affected by nM to μM levels of protons, it would be rendered insignificant by the greater operational principles. This is because the overall proton equilibriums are connected with other ions' equilibriums. There would be adsorbents/chelators on both sides that could give & take the various ions. Analogously- if we are accounting for the annual budget of a firm that runs into millions, a few transactions/donations of a few hundreds or thousands won't matter in the overall summary. Therefore, even if one were to overlook the logical catch22 we have to overcome in Mitchell's hypothesis (that of accepting the impossible premise that energy expended is recycled, by deriving a potential difference that cannot be generated in the first place!), it can be seen that a proton-based potential generation is simply unviable or untenable. Quite naturally, most resting mitochondrial potentials were experimentally determined to be $\sim$-100 mV. {When the respiring membrane potentials were detected reaching up to $\sim$200 mV (the threshold for ATP synthesis), it involved copious ROS generation. [Nicholls, 2004]. We shall return to the significance of this observation in the second part of my work.} Very importantly, accounting for the "unaccountability" of required energy with another term (a chemical potential) does not give chemiosmosis any respite. If this potential results owing to a redox equivalents' consumption leading to "harness-able" proton pumping, even then the calculations above should hold. (If the chemical potential is independent, then chemiosmosis becomes redundant because the proton pumping exercise is not what leads to ATP synthesis, and it is the other source of energy that must be important. But we know that chemiosmosis based ATP synthesis relies on protons' return to matrix. Therefore, all three concepts- ETC, proton pumps, chemiosmosis – have been effectively debunked. As I stated earlier, the fact that there are no protons to pump out makes the EPCR unviable anyway!)

**6. The disruptive mechanisms of cyanide, uncouplers and ionophores:** KCN, with a molecular mass of 65 g/mol, is lethal when consumed orally at ~1.3 mg/Kg body weight. This means that for a human weighing 50 to 100 kg, a working concentration of 20 μM cyanide is lethal (assuming that average body has an equivalent density as that of water). *In situ*, the effective concentration of cyanide can only be envisaged to go far lower than what is administered, owing to loss through retention in the gut or excretion.) Cyanide, being an asymmetric species, would be relatively less hydrophobic than oxygen, and therefore, oxygen would most probably out-compete it within the plasma membrane. Further, the $K_d$ values proffered by cyanide binding at the heme centre approaches $10^{-4}$ to $10^{-3}$ M levels for most heme (Fe-III) proteins under *in vitro* conditions, and there is little logic to imagine that in vivo conditions would drastically change these values. It must be remembered that while the anionic form binds to Fe, the pKa of HCN is 9.4, at least two units higher than physiological pH. Why then, such a highly evolved oxygen-binding machinery of a plasma-membrane embedded Complex IV loses out even to trace quantities of cyanide? Why do small amounts (of even micromolar concentrations; up to a few tens of milligrams for an average human being) of an agent like cyanide mess with the respiratory logic? Further, an uncoupler like dinitrophenol would have low mobility across the membrane, owing to the charge on its nitro groups. One wonders how or why they should keep dissipating an assumed gradient across the inner membrane. If it did (as is presumed, by shuttling protons in and out of mitochondria and thereby, messing the pumping machinery), why is it that the intact mitochondrial ETC (oxygen uptake) dependent on ADP & Pi, when it is known that uncouplers/ionophores can delink electron transfer from ATP synthesis? Seen in another perspective- mitochondrial fragments could carry out ET, but not ATP synthesis. How does EPCR come to terms with the fact that intact mitochondrial systems need ADP & Pi, when the ETC does not need them at any stage? Clearly, EPCR explanation does not have any answer! For a more detailed discussion on uncouplers and ionophores and the errors involved in the interpretation of their roles/effects, please read elaborate critiques already published by "doubters" of the prevailing EPCR [Ling, 1981; Slater 1987; Nath, 2010]. Further, please see that my group has shown that cyanide inhibition of heme enzyme activities could be mediated via alternate modalities too [Parashar et al, 2014].

**7. Rotary synthesis of ATP by Complex V?** It must be clear now that there are no protons to be pumped out and there is no ETC to facilitate that process. Even if all that happened still, the energy that is used by ETC-proton pumps for "pumping out protons" would be lost and there is no way the same can be channelized. The natural question that one could then ask is- what is being tapped by ATPase? Clearly, since the system is isothermal, the kinetic energy of the molecules of the inter-membrane milieu is no more than that of the matrix. I have established that the potential difference seen in respiring cells is not connected to ETC-proton pumps or chemiosmosis. Then, how does Complex V function as a rotary synthase? What if we indulged a hypothesis that it perhaps did not synthesize ATP in the first place? Since ETC did not exist, proton pumps did not exist, and chemiosmosis was clearly demonstrated to be a mirage, this could also be a distinct possibility. But before indulging this idea, let us concede all points to existing beliefs and accept that ATPase is a perfectly reversible rotary enzyme, that is- it works equally well as a synthase. It is now opportune that the reader avails some fundamental concepts in the area as made available in Item 6, Supplementary Information. (The content therein also affords a follow-through of some points discussed herein.)

Now, let us try to understand the equilibriums/kinetics of ATP(synth)ase and explain it by reducing it to a single catalytic site model. Let us consider the feasibility of achieving perfect reversibility, with this simple model. The $F_o$ portion is supposed to transport protons at rates approaching of $10^3$ to $10^4$ $s^{-1}$. This corresponds to a maximal ATP synthesis rate of ~$10^3$ per second, because 3-4 protons' movement across the $F_o$ module are associated with the synthesis/hydrolysis of one ATP molecule. And this value matches with literature's reports of rates ($10^1$ to $10^3$ $s^{-1}$). Therefore, it means that $F_1$ ATPsynthase must function instantaneously (with respect to the diffusion time scales) to form product(s). With micromolar level concentrations of any catalyst that works on two-electron logic, the maximal range permitted for catalysis is in the diffusion limited regime, ~$10^2$ $s^{-1}$. If you consider that protons are a reactant (at alkaline conditions) or participant (as the tripping agent to gyrate the motor) in the system, the maximum catalysis rate that can be achieved can only be ~$10^1$ $s^{-1}$. One can clearly see that such an ATPase cannot synthesize ATP using protons because of limitations imposed by proton availability. Let's still move on, granting Boyer some more consideration.

From texts, I could see that $F_1$ module of ATPase has a several order lower $K_d$ for ATP ($\sim 10^{-12}$ M!?) than the $K_d$ for ADP ($\sim 10^{-5}$ M) or Pi ($10^{-2}$ M). Let us be impartial and assume a diffusion-limited second-order on-rate of $10^8$ $M^{-1}$ $s^{-1}$ for all these entities. Then, the off-rates of $F_1$ would be $10^{-4}$ $s^{-1}$ for ATP, $10^3$ $s^{-1}$ for ADP and $10^5$ $s^{-1}$ for Pi. We can now see with a single site consideration that while the catalytic rate terms are appreciable for the hydrolysis reaction (in which the enzyme has high affinity for the substrate, ATP), it is impossible to envisage how the reverse reaction of esterification/phosphorylation could ever be feasible. This is because unlike the hydrolysis, for the esterification reaction-

(i) Two substrate molecules should bind simultaneously or bind one after the other, to adjacent locus of the $F_1$ module. Even though we have disregarded concentration and diffusion constraints, the requirement would impose significant spatio-temporal (probability) limitations, thereby lowering rates.

(ii) The $F_1$ module has much lower affinities for the two substrates ADP & Pi (than ATP) and the $K_d$ calculation shows that ADP is more likely to be unbound than be bound with the $F_1$ module.

(iii) ATP is usually found in the mitochondria at about an order higher concentration than ADP and Pi is only in the range of its $K_d$ value.

The off-rates for ADP and Pi are $10^7$ to $10^9$ folds higher than ATP off-rates, implying that it is very unlikely that this enzyme module could ever function as a synthase, under the concentration ranges that is usually found in the mitochondrion (Pi = $10^{-2}$ M, ATP = $10^{-3}$ M & ADP = $10^{-4}$ M). The only way that the enzyme can work as a hydrolyzer in one direction and esterifier in the other is by having its affinity reversed; and that is just wishful thinking! (Quite like Mitchell wanted the inner membrane to be conveniently permeable!) Another way to understand this is by looking at the way the stalk/shaft functions in a rotator mode. In the hydrolytic mode, the stalk rotates because of the ATP binding inside, leading to a hydrolysis reaction that releases energy. In this mode, an ATP molecule can bind efficiently, get hydrolyzed and this energy can be used to twirl the shaft. In the meanwhile, there is a good probability that the next ATP would have been bound on the adjacent site and the hydrolytic process goes on, and the ATPase can potentially work as a rotary hydrolyzer. But in the esterification mode, the causative is the binding of protons on the outside. A surplus of protons (in the best case scenario favoring the EPCR hypothesis) would keep rotating the the shaft and since the F1 portion has high affinity for ATP, it will always out-compete the binding process of the two substrates (ADP and Pi) at the internal binding site, thereby leading to futile cycles. As per the Boyer cycle, the ATPsynthase can only liberate pre-synthesized ATP. Else, the "ATPase" could potentially synthesize ATP with very poor efficiency when there is little ATP inside and there are saturating levels of ADP and Pi inside the matrix. And this rate cannot approach the high rates that are attributed to ATPsynthase.

Let me project the proper cause-consequence realistic perspective further. As far as we are concerned, the three sites of ATP(synth)ase would cycle through several possibilities- bound to ATP / bound to ADP + Pi / remain vacant / bound to ADP only / bound to Pi only. For the sake of simplicity and providing ample opportunity to Boyer, we shall deal with only the first three possibilities. That is- at any given instant, each one of the three sites is bound to ATP, ADP+Pi and nothing, respectively. (The third state is inconsequential for our calculation but we put it there to pay respect to the probability operator. It is not obligatory that something must be bound to each site at a given instant; particularly, when the shaft spins around at a frequency of $10^3$ to $10^4$ hertz!) As per the current hypothesis, the movement of enzyme works as a liberator of a bound substrate or product molecule. Now, let's go through the cycle. At a given instant- at one site, ATP is released; at the second site, ADP + Pi is released if it was found to be bound and at the third site, the shaft just ploughs through without sponsoring anything. The point to note is that I have said it pushed out both ATP & ADP + Pi; not ATP & ATP <u>OR</u> ADP + Pi & ADP + Pi. The movement of the γ shaft cannot work as a pushing out agent of ATP in one direction and pushing out agent of ADP+Pi in the other. (Or, stated otherwise, it cannot work as a hydrolyzer in one direction and esterifyer in another.) Also, it does not have the molecular intelligence to figure out that since it is going in a given direction, it is only supposed to synthesize. Please see that this consequence is regardless of which direction the protons flow or the shaft rotates. The conformational effects brought about in the protein would be the same. The cause-consequence correlation cannot be reversed for wishful thinking. This analysis discredits the rotational synthesis view irrevocably. (You can see it also this manner- in one way, the enzyme would cycle from high ATP affinity to low ADP affinity. In the other direction, it can still cycle only from low ADP affinity to high ATP affinity. Changing the direction does not change affinities.)

Let us see things in a quantitative manner now for the scenario we envisioned in the earlier paragraph. Assume a resting state (t0) where the matrix has 10 protons and the inter-membrane space has 10000 protons. (Let us just be gracious here!) The enzyme activity now kicks in and in a second, $10^4$ protons move into the matrix through ATP(synth)ase. Let's consider that the same amount of proton can move out within the next second (t2), through the respiratory Complexes. This would be a simple staggered arrangement and it would help us understand the scenario better. Then, the enzymatic processing (esterification/hydrolysis) at the first second, t1, would be as shown in Table 3.

Table 3a shows that with the current model, the rotary synthase can only work futile cycles, releasing or binding pre-formed substrates or products. If we must imagine this system to recycle; either through the ETC-proton pumps (t2) or through $F_o$ portion of ATPase itself (t2'), the futility only intensifies. Therefore, ATPase's "rotary synthase" role must be discarded as an unreal proposition. There is no evidence or logic to the supposition that the direction of rotation (if it rotates, that is!) dramatically alters the affinities of the enzyme. A realistic scenario in which the rotary ATPase would catalytically "work" is given in Table 3b. Please note that once again, in t2', it did not matter which direction the protons flowed through $F_o$ portion. This is because ATP hydrolysis is the horse that could potentially drive the cart of protons in a given direction; not the other way around!

As per Boyer, energy is required (which was presumed to come from proton movement- a supposition already debunked) for three things- moving the shaft, detaching the ATP and getting ADP and Pi to have greater affinities for the enzyme. One finds the last postulate impossible to comprehend. I thought the shaft moved and detached anything that bound, as it moved. How can energy be used to increase binding affinity of ADP and Pi? Isn't that an "over-personalizing wish"? These are just unsubstantiated proposals that prevail only because Mitchell's chemiosmosis was around. The quaint part of it all is that purportedly, the actual synthesis step of ADP and Pi coming together to form ATP does not apparently require any "energy" on the enzyme surface! (This means that activation energy is lowered to approach a zero value.) The process is supposedly (freely) reversible on the enzyme surface. If that were so, why did the cellular system go through the whole drama of ETC, proton pumps and rotational synthesis? Homeostasis (maintenance of cellular metabolic equilibrium) could have been achieved in a much simpler way without the necessity of all the complexities. Thus, there is no way that the chemiosmosis based proton pumping could support a rotary synthesis, which is untenable even otherwise. [Please refer Item 7 for some relevant conclusions in this regard.]

**Conclusions**

EPCR hypothesis had no citadel or edifice that I had originally set out to demolish. It was a mere façade, which is razed by the queries, analyses and critical insights offered above. Non-existent protons were moved in and out to weave a fabric of unsubstantiated explanations, which had little bases in reality. It must be remembered that qualitative reasoning that captivates creative/imaginative thought faculty of man is inadequate for achieving the status of a theory. Scientific reality is grounded in quantitative justification (that is- it must answer to critical/doubtful thoughts and meticulous accounting based on sound logic). In the scientific validation process (and ultimately, concrete progression towards understanding a phenomenon), it is more acceptable to disprove a concept and that purpose is achieved with this manuscript. I hope this communication dislodges the theory status afforded to EPCR hypothesis and textbooks are corrected to the effect, at the earliest possible time. We do not need to leave behind arbitrary beliefs in science texts for our future generations to mock us. In the second part of this work [Manoj, 2017], I have elaborated upon a probable explanation for mitochondrial oxidative phosphorylation; and I request the reader to kindly peruse the same.

**TABLES & FIGURES**

**Table 1:** *A comprehensive account of the purported ETC* scheme for the reduction of one molecule of oxygen (by the circuit of Complex I – Complex II – 2 Complex III – Complex IV).

| Element | Partici pants | Steps (2e+1e) | Overall distance (highest) | Distance per electron | Overall gradient [start (low, high) end] | Unfav orable steps | Non-"route" redox centers |
|---|---|---|---|---|---|---|---|
| **Comp I** | 10 | 1 + 16 | 214 (16.9, 14) | 108 | -320 (-480, -150) +113 | 14 | 2 (FeS: N1a = -233 & N7 = -314) |
| **Comp II** | 6 | 1 + 8 | 105 (16, 11.9) | 56 | -31 (-260, +60) +113 | 4 | 1 (heme: -185) |
| **Comp III** | 6 | 0 + 12 (includin g 6 for CoQ recycle) | [170 (34 to 20) (including 100 for CoQ recycle)] x 2 | 35 (not including >50 for CoQ recycle) | +113 (-90, +300  ) +254 | 6 | Nil |
| **Comp IV** | 6 | 0 + 16 | 120 (16) | 30 | +254 (+240, +320) +820 | 8 | Nil |
| **Overall** | 24 | 54 | >750 (at least ten transactio ns are above 12) | ~230* | ~ -400 to ~ +800 (nine transactions unfavourable ) | 32 | 3 |

* This is a highly conservative estimate by any means. The value for the conservative distance of a single electron travel within "the highly efficient" supercomplex (formed by Complex I – Complex III – Complex IV) would be minimally ~350 to 400 Å [[as can be seen in Figure 1c, in the review by Enriquez and Lenaz (2014) or Figure 7b  of the review by Kuhlbrandt (2015). So, a multi-disrupted four-electron travel even in this "optimized" but "un-insulated discontinuously wired" system would total a distance ~ 1500 Å!

**Table 2: *Transient potentials*** (and their dependence on percentage of internal protons pumped out) that could be generated by concerted efforts of proton pumps in the mitochondria, assuming diverse initial states of pH. *Key values relevant to discussion are highlighted in bold.*

| No. | $[H^+]_i$ (nM) | $[H^+]_o$ (nM) | Resting potential (mV) | Transient potential (mV) | | | | | | | | |
|---|---|---|---|---|---|---|---|---|---|---|---|---|
| | | | t1 (0) | t2 (1) | t2 (10) | t2 (25) | t2 (50) | t2 (75) | t2 (90) | t2 (99) | t2 (99.9) |
| **1** | 10 | 10 | 0 | 1 | 5 | 14 | 29 | 52 | 78 | 140 | **201** |
| **2** | 10 | 50 | 43 | 43 | 46 | 52 | 64 | 83 | 108 | 169 | 230 |
| **3** | 10 | 100 | 61 | 61 | 64 | 69 | 81 | 100 | 124 | **186** | 247 |
| **4** | 10 | 500 | 104 | 104 | 106 | 111 | 122 | 141 | 165 | 226 | 287 |
| **5** | **10** | **1000** | **122** | 122 | 125 | 130 | 140 | 159 | **183** | 244 | 305 |
| **6** | 50 | 10 | -43 | -41 | -29 | -14 | 9 | 35 | 64 | 127 | 188 |
| **7** | 100 | 10 | -61 | -58 | -40 | -20 | 5 | 32 | 61 | 124 | 186 |
| **8** | 500 | 10 | -104 | -93 | -53 | -27 | 1 | 30 | 59 | 122 | 183 |
| **9** | **1000** | **10** | **-122** | -103 | -56 | -28 | 1 | 29 | **59** | 122 | **183** |
| **10** | 100 | 100 | 0 | 1 | 5 | 14 | 29 | 52 | 78 | 140 | 201 |

**Table 3 a & b: *Simple simulations of a. ATPsynthase & b. ATPase***

a.

| Time | Matrix protons | Protons outside | ATPsynthase revolutions | ATP auto-off | ADP auto-off | Uncatalyzed enzymatic ATP liberation | Uncatalyzed enzymatic ADP+Pi liberation | Net ATP synthesized/ hydrolyzed |
|---|---|---|---|---|---|---|---|---|
| **t0** | 10 | 10010 | na | na | na | na | na | na |
| **t1** | 10010 | 10 | ±1000 | 0.0001 | 1000 | 3333 | 2333 | 0 |
| **t2** | 10 | 10010 | na | 0.0001 | 1000 | 0 | 0 | 0 |
| **t2'** | 10 | 10010 | ±1000 | 0.0001 | 1000 | 3333 | 2333 | 0 |

b.

| Time | Matrix protons | Protons outside | ATPase revolutions | ATP auto-off | ADP auto-off | Enzyme-catalyzed ADP+Pi liberation | Uncatalyzed enzymatic ADP+Pi liberation | Net ATP hydrolyzed |
|---|---|---|---|---|---|---|---|---|
| **t0** | 10 | 10010 | na | na | na | na | na | na |
| **t1** | 10010 | 10 | ±1000 | 0.0001 | 1000 | 3333 | 2333 | 3333 |
| **t2** | 10 | 10010 | na | 0.0001 | 1000 | 0 | 0 | 0 |
| **t2'** | 10 | 10010 | ±1000 | 0.0001 | 1000 | 3333 | 2333 | 3333 |

**Figure 1:** *A representation of the minimal aspects of the protein structures and the salient events therein.* (Adapted unabashedly from Sazanov, 2015) The four complexes on the left side constitute the ETC whereas the lone complex on the right constitutes the ATP synthesizing molecular motor. However, the sites of proton pumps (Complexes I, III & IV) are functionally equivalent with ATP synthesis sites too. IMS & IPLM stand for inter-membrane space and inner phospholipid membrane respectively. As per the current perception, 9 - 12 protons are required for one rotation of the ATP synthase, concomitantly yielding three molecules of ATP.

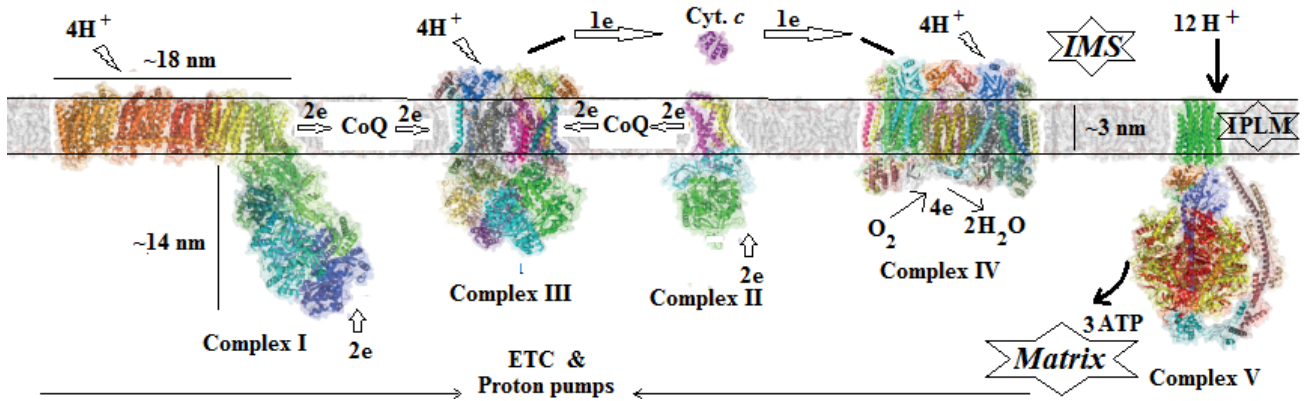

**Figure 2:** *A simple spatio-temporal model of electron transport cum proton pumping coupled oxidative phosphorylation constructed as per the prevailing hypothesis.* The arrows point to the movement of protons in time. The two modules of ETC+proton pump and chemiosmotic ATP synthesis are represented by g and h respectively.

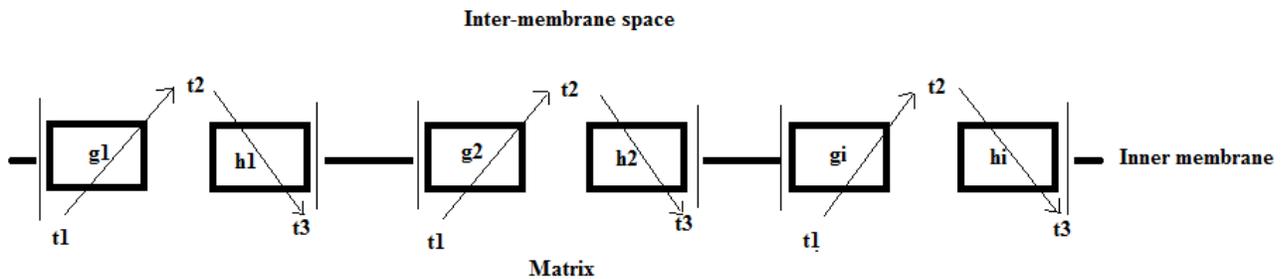

# Supplementary information: *Mitochondrial oxidative phosphorylation…*

## *Kelath Murali Manoj (2017)*

### Item 1: My long-standing doubts regarding EPCR

As a disenchanted biology graduate (when I had much greater interest in appreciating the morphological and behavioral aspects of biology and cared less for its molecular foundations), I found it difficult to assimilate the essential components of mOxPhos, as prescribed in the syllabus for my Masters Program. Here is what I found difficult to digest then (in 1990, while browsing through renowned textbooks authored by the likes of Albert Lehninger & Lupert Stryer)-

*Questioning the theoretical foundations:* What is so special with the inner mitochondrial membrane that it alone is impermeable to protons? If it is impermeable to protons and if the inter-membrane aqueous phase is fully (physically and functionally) demarcated from the mitochondrial matrix, how is the osmotic mechanism operational? Osmotic equilibration forces could work towards attaining equilibrium across a semipermeable membrane but this equilibration process would require "a decent connectivity" across the two phases. An appropriate analogy that one could quote is- In a hydroelectric power plant, the generator's turbines would only rotate if a large volume of water rushed through (by opening the gates at the catchment on top), and the water that flows across the turbines should also have a free exit at the bottom. If the water trickled through a crack on the wall to stagnate at the bottom, the turbine would rather get rusted, than rotate! For the Complexes-mediated ejection of protons into the inter-membrane space (as a result of ETC), if we were to assume that the outer membrane is freely permeable to protons (to enhance the spontaneity), how could it serve to retain the gradient for the ATP synthesis part of the deal? Further, even in terms of semantics/definition, osmosis would entail the movement of water (the solvent) molecules across the semipermeable membrane, not protons! If it was the proton alone that spontaneously moved in as a "powerful" process, it would de-facto violate the "inner membrane semipermeability" assumption we started off with! Most critically, the proton-pumping process seemed to be chemically disconnected (or, very remotely connected) with the ADP phosphorylation process in the mitochondrial matrix. This predicament just did not satisfy the chemical logic of the whole cycle.

*Concerns on experimental findings and energetics:* How could a few protons trickling down the transmembrane part of the multiprotein-complex of $F_oF_1H^+$-ATPase physically serve as the *tour de force* accomplishing ATP synthesis? It seemed too insignificant a drive. Though I did not calculate it then, I had this gut feeling! I compared the Complex V setup to a medieval hand-driven (purely mechanical) stone grain-grinder. Whole grains introduced from the top into a small hole (analogous to protons' inward movement) leads to the grains entering a gap between two flat stones. The grains are ground between the two stones to give powdered flour (analogous to ATP formation) by mechanical movement of the stone at the top (enabled by the operator's application of a torque). The operational principle herein is that the weight of the top stone works (gravity, friction, torque, etc. all combined!) on the grains held against the stationary bottom. In comparison here, it is difficult to envisage that the mechano-chemical changes brought about by the inward trickling of 3-4 protons (I could not see this process as a chemical reaction, electrical field or a mechanical push either!) would afford enough torsion to move the stalk (γ) through the αβ bulb. Even if the huge excess of pumps met with success in their vitally deterministic mission (that is- let us suppose fallaciously that they found and pumped every single one of the protons out and created very high potential across the small mitochondrial membrane owing to a disparity in these small amounts of protons), I felt that it cannot perform a major work across the same membrane because the amount of charge would be too little. But was it the charge or the numbers+mass or all of it combined that was doing the trick? How does the contraption work? We now know that $> 10^3$ protons are required for a single rotation of the bacterial flagellum. Here, we have only $\leq 10^1$ protons achieving a single rotation of the ATPsynthase!

The chemiosmosis hypothesis apparently seemed to afford a qualitative logic for the requirement of intact membrane and oxygen cum reduced substrate for ATP synthesis. It also went well with the finding that mitochondrial fragments could carry out the ETC process and "uncouplers and ionophores" could delink electron transfers (ET) from ATP synthesis. However, I could not see why intact mitochondrial ETC (oxygen uptake) is dependent on ADP & Pi. On the same note, why should blockers of ATP synthesis (Venturicidin and Oligomycin; membrane pore/channel formers) inhibit intact mitochondrial ETC (oxygen uptake)? Key experimental findings reported in the field did not seem to agree with each other. When proton motive force alone was found to be inadequate to provide for the energetics, it was presumed that difference in ionic concentrations across the membrane could also lead to ATP formation. Now, I could never fathom how the molecular motor in ATP synthase could "sense" this electrical energy or concentration difference and tap into this drive for a biological synthesis. I could also not find any pointers for the source of replenishment/resurrection of the electric potential that would be dissipated with the inward proton movement/coupling event. (Does the membrane potential not get dissipated at all? In that case, where from comes the ultimate source of energy? Was this some kind of vital perpetual machine?)

*Evolutionary perspectives:* The prevailing explanation smirked at me, reeking of a creationist undercurrent. The wiring logic of the various ETC complexes within the membrane eluded me because it sought an external and primordial "ordering" agent for any life sustaining energy formation process (and predestined routes of electron transfer even for freely soluble and diffusible entities). Also, why should the ubiquitous ubiquinone serve as an electron relay that gets two electrons only from Complex I or II and lead to only Compound III? What drives such "witless" molecules to be so directional in their motion and electron transfer? How can the low values of proton concentration and permeability/mobility within the plasma membrane sustain the high rates sought of ETC with the hopelessly circuitous mechanisms for CoQ/Complex III interaction? If the purpose of ETC was to have protons pumped out across the inner mitochondrial membrane, why did the Complexes I and II not evolve to achieve the task on their own? Why were the other elements of ETC featured within this vague scheme? How does protein-protein binding mediated electron transfer (followed by long-distance electron tunneling) get to be so directional and efficient? How could the bulky proteins defy diffusion constraints within the highly viscous and low-energy regime of membrane lipids and inter-membrane spaces? The whole setup appeared to be highly inefficient and superfluous, at the same instant. Why should the spontaneous process of substrate oxidation lead to an uphill trans-membrane pumping of protons? *What could have been the evolutionary pressures or stochastic phenomena that could have brought about this "gambit" causality/effect correlation?* How could such a highly ordered and yet indirect process, followed by a highly deterministic "coupling with a sophisticated motor" serve as the primordial/quintessential pivot for the evolution of oxygen-centered life? ATP can also be synthesized by alternate routes- for example, by the phosphoryl group transfer in substrate-level phosphorylation. The sophisticated "proton gradient driven molecular motor" option is not obligatory. Then, why should life process evolve to use such a superfluous, complicated and wasteful/hazardous system (that is known to generate ROS)?

## Item 2: A critical commentary on CoQ Cycle

The CoQ cycle (can be deemed a control point!) and its execution at Complex III happens to be a highly ingenious mechanism proposed in modern biochemistry. As per erstwhile ideas, three different molecular species are supposed to simultaneously (or at some precisely defined time points) bind to three different sites of Complex III. In the first step of the cycle, two intra-membrane sites Qo and Qi bind to $CoQH_2$ and CoQ respectively and an inter-membrane (Cyt. c1) site binds to oxidized Cyt. c. This (supposedly!) leads to an expulsion of 2 protons to inter-membrane space and reduction of Cyt. c and partial reduction of CoQ at Qi site (leading to the semiquinone). In the second step, a molecule of CoQ and a molecule of oxidized Cyt. c bind at Qo and Cyt c1 respective sites (and the Qi site remains occupied with the one-electron reduced, semiquinone form of CoQ). Subsequently, Complex III draws two protons from the matrix (or is assisted by Complex II in this regard?), and gives the formation/release of reduced CoQH2 and Cyt. c at sites Qi and Cyt c1 respectively (with concomitant release of CoQ at Qo). The reactions could be summated as follows-

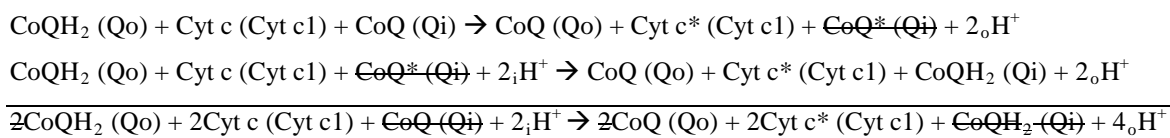

$CoQH_2$ (Qo) + Cyt c (Cyt c1) + CoQ (Qi) → CoQ (Qo) + Cyt c* (Cyt c1) + ~~CoQ* (Qi)~~ + $2_oH^+$

$CoQH_2$ (Qo) + Cyt c (Cyt c1) + ~~CoQ* (Qi)~~ + $2_iH^+$ → CoQ (Qo) + Cyt c* (Cyt c1) + $CoQH_2$ (Qi) + $2_oH^+$

~~2~~$CoQH_2$ (Qo) + 2Cyt c (Cyt c1) + ~~CoQ (Qi)~~ + $2_iH^+$ → ~~2~~CoQ (Qo) + 2Cyt c* (Cyt c1) + ~~$CoQH_2$ (Qi)~~ + $4_oH^+$

The two protons taken up from the matrix side (shown here in step 2 only) could also be achieved via two steps (of one proton each, in step 1 and 2). The scheme, though elegant, is highly fastidious and quite difficult to justify practically. One wonders about the efficiency of the system at one hand and the thermodynamic drive at the other! Why or how would the electrons flow from $CoQH_2$ back to CoQ, when there are plenty of other options available? Why should the quinone intermediate(s) give and take electrons in the same cycle, at the same Complex?

While it seems possible, how does this retain directionality with respect to the overall thermodynamic drive involved? How is the molecular intelligence of this fantastic scheme maintained? While the electronics can be easily written down and the overall scheme drawn, it would have very little probability to function in the real arena. And let me reiterate that anytime this fastidious scheme does not work, the whole ETC circuitry breaks down!

**Item 3: Some oxygen-based species that could be encountered in mitochondria**

**Table 1:** Electronic distribution and naming conventions of select/relevant oxygen-centered small entities

| Species | Notation | Stature (charge) & relevant pKa | Electronic distribution within the orbitals | | | | | | |
|---|---|---|---|---|---|---|---|---|---|
| **Oxygen** | O | Atom (0) | ↑↓ | ↑↓ | ↑↓ | | | | |
| **Triplet Oxygen** | $^3O_2$ | Diatomic diradical molecule (0). (Normal dioxygen) | ↑↓ | ↑↓ | ↑↓ | ↑↓ | ↑↓ | ↑ | ↑ |
| **Singlet Oxygen** | $^1O_2$ | Diatomic Molecule (0) (Excited dioxygen; more reactive to organics than singlet oxygen.) | ↑↓ | ↑↓ | ↑↓ | ↑↓ | ↑↓ | ↑↓ | |
| **Superoxide** | $O_2^{*-}$ | Diatomic radical ion (-); becomes triatomic perhydroxyl/hydroperoxyl uncharged radical on protonation at pH 4.8. (Protonated species more reactive to organics than superoxide ion.) | ↑↓ | ↑↓ | ↑↓ | ↑↓ | ↑↓ | ↑↓ | ↑ |
| **Hydrogen peroxide** | $H_2O_2$ | Tetra-atomic molecule (0); becomes triatomic hydroperoxide ion on deprotonation at 11.7. | ↑↓ | ↑↓ | ↑↓ | ↑↓ | ↑↓ | ↑↓ | ↑↓ |
| **Hydroxyl** | OH* | Di(hetero)atomic uncharged radical (0), 11.5. Deprotonated species less reactive. | ↑↓ | ↑↓ | ↑↓ | ↑ | | | |
| **Hydroxide** | OH⁻ | Di(hetero)atomic ion (-); becomes triatomic water molecule on protonation. | ↑↓ | ↑↓ | ↑↓ | ↑↓ | | | |
| **Oxide** | $O^{2-}$ | Monoatomic ion (--) | ↑↓ | ↑↓ | ↑↓ | ↑↓ | | | |
| **Water** | $H_2O$ | Triatomic molecule (0), 14. | ↑↓ | ↑↓ | ↑↓ | ↑↓ | | | |

**Item 4: Detailed analysis of the purported mitochondrial ETC**

We can peruse the data available for each one of the complexes within this circuitry. [[In the sections to follow, I request that the reader kindly refers the original source of data, verify the same and quote that citation, and not this work as the source of values of redox potentials!]] I present the ETC for one molecule of water reduction through the given scheme of EPCR, with the following modularity [Comp I : Comp II : 2 Comp III : Comp IV].

Complex I supposedly takes two electrons from NADH and passes it on to CoQ. The currently perceived ETC route within the matrix-protruding arm of Complex I is given in Table 1 below:

**Table 1**: Analysis of electron transport chain within Complex I. NADH binding was found to approach a value of $10^8$ $M^{-1}s^{-1}$ (corresponding to 0.2 µs binding time). (Data for this table are derived from the following publications- Treberg & Brand, 2011; Verkhovskaya et al, 2008; Sazanov, 2015; Bridges et al, 2012; Medvedev et al, 2010; Moser et al, 2006) NADH binding rate was found to approach a value of $10^8$ $M^{-1}s^{-1}$ (corresponding to 0.2 µs binding time- fast, approaching diffusion limitations!). Further, Mossbauer analysis showed that some FeS centres (N4 & N6a) along the route were not reduced [Bridges et al, 2012]. As can be seen (from the table above), these loci represent transfer zones of unfavorable distances and/or redox gradient.

| Step | Donor (mV) | Acceptor / Product (mV) | Distance (Å) | Favorable ΔE? | Favorable Δd? | Electron(s) |
|---|---|---|---|---|---|---|
| 1 | NADH (-320) | FMN (-340 mV) | <3? | Probable; with relatively higher NADH | Probable | 2e / step (proton required) |
| 2 | FMNH$_2$ (-340 2e process; -389 and -293 respectively for 1$^{st}$ and 2$^{nd}$ steps) | N3 (-250 to -321); N1a (-233) | 10.9 (7.6); 13.5 (12.3) | Probable; Probable | Probable; Less probable (no further connectivity) | 1e / step (x 2) |
| 3 | N3 (-250) | N1b (-240 to -420) | 14.2 (11) | Less probable | Less probable | 1e / step (x 2) |

| Step | Donor | Acceptor / Product | Distance | Favorable ΔE? | Favorable Δd? | Electron(s) |
|---|---|---|---|---|---|---|
| **4** | N1b (-370) | N4 (-250 to -291) | 13.9 (10.7) | Probable | Less probable | 1e / step (x 2) |
| **5*** | N4 (-250) | N5 (-270 to -480); N7 (-314) | 12.2 (8.5); 24.2 (20.5) | Less Probable; Less probable | Probable; Less probable | 1e / step (x 2) |
| **6** | N5 (-430) | N6a (-250 to -325) | 16.9 (14) | Probable | Less probable | 1e / step (x 2) |
| **7*** | N6a (-250) | N6b (-188 to -420) | 12.2 (9.4) | Less probable | Probable | 1e / step (x 2) |
| **8** | N6b (-420) | N2 (-50 to -200; consensus = -150) | 14.2 (10.5) | Probable | Less probable | 1e / step (x 2) |
| **9** | N2 (-150) | CoQ (-300 to -120) | 11.9 (8.6) | Less probable | Probable | 1e / step (x 2) |

Next, Complex II is the enzyme that takes electrons from succinate to give it on to CoQ. The analysis of its redox centers is given in Table 2.

**Table 2**: ETC route in Complex II. (Data for this table was taken from Yankovskaya et al, 2003; Sun et al, 2005; Anderson et al, 2005) The turnover rates of the final step of CoQ reduction was found to be $10^2$ to $10^3$ $s^{-1}$.

| Step | Donor (mV) | Acceptor / Product (mV) | Distance (Å) | Favorable ΔE? | Favorable Δd? | Electron(s) |
|---|---|---|---|---|---|---|
| **1** | Succinate (-31) | FADH2 (-79) | 4.6 (2.5) | Probable | Probable | 2e / step |
| **2** | FADH2 (-79) | 2Fe-2S (0) | 16 (11.1) | Probable | Low probability | 1e / step (x 2) |
| **3** | 2Fe-2S (0) | 4Fe-4S (-260) | 12.4 (9.3) | Low probability | Low probability | 1e / step (x 2) |
| **4** | 4Fe-4S (-260) | 3Fe-4S (+60) | 11.9 (9.1) | Probable | Probable | 1e / step (x 2) |
| **5a** | 3Fe-4S | CoQ (+113) | 11 (7.6) | Probable | Probable | 1e / step (x |
| | (+60) | | | | | 2) |
| **5b** | 3Fe-4S (+60) | Heme b (-185) | 18.5 (11.4) | Low probability | Low probability | 1e / step (x 2) |
| **6** | Heme b (-185) | CoQ (+113) | 6.5 (9.8) | Probable | Probable | 1e / step (x 2) |

The analysis of the redox centres of Complex III comes next and the data is given in Table 3.

**Table 3**: Redox centres of Complex III. (Source- Crofts webpage at UIUC[1]; Iwata et al, 1998 & Zhang et al, 1998) The distances between the respective hemes and FeS centres in the dimers are distances from 21 to 63 Angstroms apart and therefore, they are not considered relevant for intermolecular ET phenomena. The data analysis is stemmed on the belief system that ubiquinone (reduced and oxidized) interact at two locale on the enzyme- the reduced species interacts via the Fe-S protein on the intermembrane side and the oxidized species interacts via the Cyt bH on the matrix side. The FeS Rieske protein is a bifurcating point, with options to give the electron to Cyt c1 in the intermembrane space or Cyt bL towards the matrix side of the inner membrane. The purported electron flow is is more complex than what is given herein.[2] The last step electron transfer timescale (to Cyt c) is in the range of $10^0$ to $10^2$ μs (the fastest step of the overall process) and each of the other steps may incur a time window of $10^2$ to $10^3$ μs.

| Step | Donor (mV) | Acceptor / Product (mV) | Distance (Å) | Favorable ΔE? | Favorable Δd? | Electron(s) |
|---|---|---|---|---|---|---|
| **1** | CoQH₂ (+45 to +113) | FeS (+300) | >7? (two phases) | Probable | Seems OK | 1e / step (x 2) |
| **2a** | FeS (+300) | Cyt c1 (+270) | 21.3 to 31.6 | Low probability | Low probability | 1e / step (x 2) |
| **2b** | FeS (+300) | Cyt bL (-90) | 34.3 to 26.4 | Low probability | Low probability | 1e / step (x 2) |
| **3a** | Cyt.c1 (+270) | Cyt c (+254) | <4? | Probable | Probable | 1e / step (x 2) |
| **3b** | Cyt bL (-90) | Cyt bH (+50) | 20.5 (13) | Probable | Low probability | 1e / step (x 2) |
| **4** | Cyt bH | CoQ (+90) | <4? | Probable | Probable (?) | 1e / step (x |

---

[1] For a quick reference- ***http://www.life.illinois.edu/crofts/bc-complex_site/***
[2] But there is no point in going to all the details here, anyway!

| | | | | |
|---|---|---|---|---|
| (+50) | | | | 2) |

Finally, we analyze Complex IV (the final destination of ETC), the supposed site of water formation, in Table 4. (Seriously considering the loyalty of oxygen dedicatedly resting at this station, waiting for the four electrons and protons to be sequentially acquired from Cyt. c- it reminds me the marvelous comical fantasy of Tony Stark transforming into a la Iron-Man through a programmed addition of several peripheral body suit/armor components!)

**Table 4**: The ETC in Complex IV. (Source Moser et al, 2006 and hyperphysics webpage[3])

| Step | Donor (mV) | Acceptor / Product (mV) | Distance (Å) | Favorable ΔE? | Favorable Δd? | Electron(s) |
|---|---|---|---|---|---|---|
| **1** | Cyt c* (+254) | CuA (+240) | <10 | Less probable | probable | 1e / step (x4) |
| **2** | CuA (+240) | Heme A (+260) | 16.1 | Probable | Less probability | 1e / step (x4) |
| **3** | Heme A (+260) | Heme A3-CuB (+280 to +320) | 7 to 12 | Probable | Probable | 1e / step (x4) |
| **4** | Heme A3-CuB (+280 to +320) | $O_2$ (+820) | <3? | Probable | Probable | 1e / step (x4) |

**Item 5: A circuitous unlearning of the fallacy of chemiosmosis**

**a.** *Analogies with the waterfall and watermill to point out the "flaw" (pun intended!) in logic:* Now, we come to the crux of this writing. Here, I shall call the bluff on the prevailing understanding. Since several seasoned scientists and specialists have bought into the "mirage" of "chemiosmosis" (I can afford to call it so, since I have already debunked ETC-proton pump concepts, on which chemiosmosis rested!) over half a century now, I am left with no other choice than to campaign to the uninitiated. Further, I must go through a rather circuitous route (quite like the ETC!) so that the decades' old futile explanations are shown the door out, in a very ceremonious way. I shall now try to convince you that chemiosmosis is an "unscientific" bunch of ideas.

---

[3] Quick reference: ***http://hyperphysics.phy-astr.gsu.edu/hbase/Chemical/redoxp.html***

In terms of logic, chemiosmosis is analogous to Maurits Cornelius Escher's much famous work, "The Waterfall" (and the outcome is akin to his work- "Ants on a Moebius Loop"). "The Waterfall" drawn by Emil Escher happens to be one of my favorite works of art, clearly pointing out how perceptions can be deceptive. The fabulous work is reproduced here as Figure 1a. By making a few impossible connections and drawing with a few inappropriate angles, the water always seems to flow downhill at any particular locus in the image, no matter where! This scenario is what the EPCR hypothesis affords us. Quite simply, once Mitchell's idea was sold to the world, the other ideas just coalesced around it. Once people bought into this Moebius coin, anytime it was tossed and called, the only side that it had was what it landed on! Once again, I would like to peruse Escher's excellent work, Figure 1b, to highlight the effect. The ants on the Moebius loop only experience the dimension of their given plane. (They may believe that they are getting somewhere traversing on the loop, but they don't and that is because they can't!) By assuming unreal postulates as founding assumptions, and thereafter, employing flawed protocols to reaffirm the faulty assumptions, the fabric of logic and reality were broken and resealed.[4]

**Figure 1a & 1b**: Escher's creative visual treats on bending and mixing spatio-temporal dimensions

---

[4] Another analogy, this time with numbers, to give you a picture of what happened: 7 guys hailed a cab and the fare rand to 28 bucks. Now, a mathematical genius in one of them convinced the rest that each one of them had to contribute 13 bucks apiece, with the following protocol for division. 28/7 = 13. How? 7 goes once in 8. So, write 1 on top. Then, 7 is subtracted from 8 to get 1, which is written down. Now, the remaining 2 is brought down to the left of 1, to get 21 as the remainder. Next step- 7 goes thrice in 21 and therefore, writes 3 down next to the earlier derived 1 and you have the exact number of 13! A cynic among the bunch felt that there was something wrong in the calculation and 13 seemed too big! He got out a piece of paper and said that if this calculation was true, then 13 added up 7 times should give the same number of 28. He wrote the number 13 seven times, one below the other. Now, the addition was carried out in the following manner:  7+7+7+7+7+7+7 = 21; 1+1+1+1+1+1+1 = 7; therefore, 21+7 = 28. *Quod Erat Demonstrandum!*

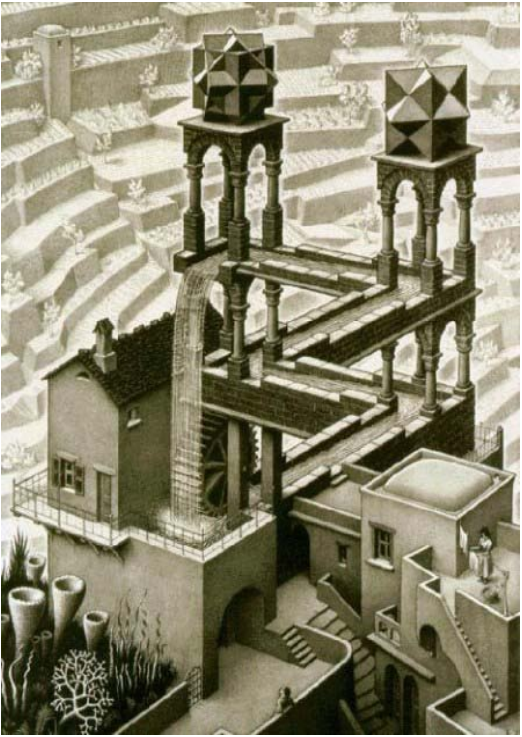
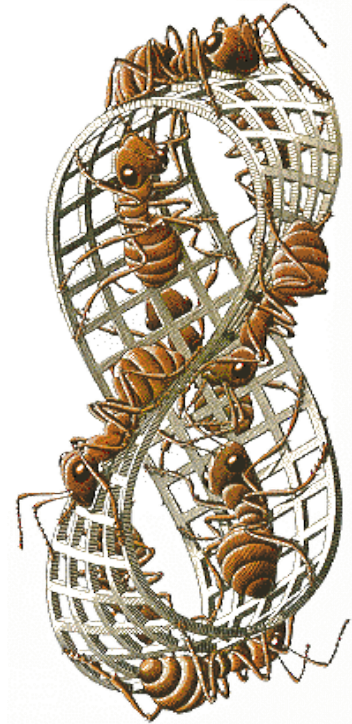

Let us hypothesize that the semipermeable ("impermeable to protons") inner membrane of mitochondria serves as a molecular machine that does the useful work of synthesizing energy-rich ATP molecules using the free energy available from the oxidation of reducing equivalents from NADH/succinate. I think that anyone would buy into this hypothesis/analogy, given our current status of awareness.

Now, I would invite you to make an analogy between the module of mitochondrial membrane (that serves as the proton transfer conduit) and a watermill (that serves as water conduit). This is a simple exercise because the watermill is also a simple machine that uses the free energy (the potential energy of water) to do the useful work of driving a pump or serving as electricity generator. Most would think that this analogy is appropriate. But soon enough, we shall note that it is not that simple! (We shall find that there are quite a few logical catch 22s and zugzwangs in this assumption!)

The three fundamental and necessary attributes of the watermill (working machine) are-
(i) free energy should be expended spontaneously (item 1),

(ii) item 1 enables the mill to perform useful work (item 2) and

(iii) item 2 justifies "man's benefit-seeking perspective" of getting some useful work done, albeit at an operating cost (item 3).

The three attributes are interconnected and each step must be relatively accountable and must also abide by the fundamental laws of physics. We can see that a single unidirectional flow of water through the mill achieves all the three attributes of the machine.

Now, let us assume the watermill analogy scenario when protons are taken from the matrix to be pumped outward to the inter-membrane space by the respiratory complexes. Starting at time $t_1$, in the **after-bay located at the bottom (in the matrix),** energy is expended to pump the **water (protons) into the fore-bay positioned upwards (inter-membrane space)**, through a locus in space (let's call it module $g$); and let us grant it that by the end of this process, the time has become $t_2$. Let's see if our machine satisfies the three requirements of the analogy. The membrane spontaneously expended free energy available from oxidation of NADH/succinate in the matrix (item 1), to pump protons out into the inter-membrane phase (item 2). Is this the useful work that we required (item 3)? One can't really say, because the protons are not useful just because they are pumped out, and therefore, the analogy does not hold yet!

Let's take the scenario a bit further. In a second step (starting at say, time $t_2$), the **water (protons)** from **the fore-bay (inter-membrane space)** is comes back into the **after-bay (matrix)**, through yet another locus in space (let's call it module $h$), where we made it go through an electricity generator (analogous to ATP synthase). Let's say that the time after this exercise is $t_3$ and let's accept it that the "machine" has gone through one full cycle (and that the overall process is repeatable).

Please see that it is only because the second step was continued after the first step, we have retained the terminologies of fore-bay and after-bay (and this is just a semantic connectivity). Here is the important catch- Since the membrane is impermeable to protons, there are no equilibrium seeking forces that the membrane could tap. Therefore, the second step could not be spontaneous and would have to consume energy! This statement would be true no matter how efficiently one arranges the vectorial transport in the membrane. Say- the membrane pores/pumps are funnel shaped, with a narrow entry at the matrix side and wide exit at the inter-membrane space. Then, the return into the matrix would be relatively more facile! But then, we would end up going against the fundamental assumption that the membrane is impermeable to protons, a necessary criterion to create a gradient! Even if we change the definition from impermeable to directionally permeable, you will see that it does not help the cause in the context.

Once again, let's see if our membrane machine satisfies the terms of analogy. In the current scenario, the membrane expended energy (this is not derived by NADH oxidation in the matrix, but avails another independent source) by moving protons from inter-membrane space into the matrix (item 1). Is this what we required of the membrane machine (item 3)? We cannot really say until we find what work was done by the water (protons) at the generator or pump (item 2). If the generator did not yield any electricity, then items 2 or 3 are not justified and we do not have a machine. On the other hand, if the generator/pump yielded more electricity/work than permitted by the laws of physics, then we have been sold an impossible machine.[5] But by now, some of you would have noted that the second step had no functional obligation/connection with the first step. Therefore, the first step was a mere waste and as a consequence, even without knowing the outcome of item 2, we can safely conclude that inner mitochondrial membrane could not serve as an effective machine, with the EPCR hypothesis![6]

Did you note the extra whorls that the analogy seeks for its justification and how unreal assumptions were woven into the chemiosmosis hypothesis and conveniently annulled in between the steps, to get us to merge the first and second step? The membrane machine allows two movements across itself and in both these moves, the spontaneity is only ensured at the expense of two independent sources of energy! In the first one, energy is expended by fuel oxidation and in the second one, a pre-existing potential[7] is supposed to serve the purpose[8]. If the second step depended on the energy derived from the first, things would be different and we could have a straight and simple analogy with the watermill! Now (notwithstanding that we do not have any information whether the outward and inward proton moevements did any useful work), a logical exercise is to ask if you would use such a watermill scenario where you have to pump water up and then pump water down and make that water do work when it comes down. Clearly, this is not a way of getting work done spontaneously in real life![9] Why? Because we are doing relatively more work to get a lesser work done; or we have little tools at our disposal and therefore, we are destined to employ a very wasteful machine![10]

---

[5] Or, rephrased- we have been given an emperor's silk gown by some seamsters!

[6] If a straight line can be expressed with y = mx + c, people from sound scientific background do not engineer a straight line with the incorporation of any more (redundant) variables into the equation!

[7] This concept is so complex that it is beyond the scope of discussion in the current work.

Regardless of the discussion in the earlier two paragraphs, let's do a stock-taking of the process till now. The membrane machine requires **at least two distinct steps** involving **two different modules**. (The first one was supposed to use a bigger original source of free energy and the second one was originally sold to us as a transiently generated quantum that resulted to because of expending the original bigger source of energy in the first step! We have now debunked this falsity.) We know that **the two modules (g and h) must occupy two non-overlapping loci in space** (*and these two loci must not be the same as the matrix and/or inter-membrane loci either!*) and **the two events must transpire through three non-exchangeable and non-overlapping points in time**. Further, **the h module would need to do some useful work** for the inner membrane could qualify for a machine! These summations shall enable us to assess the energy used at g and work done at h modules, which shall be discussed in a later section.

---

[8] At the second step, if protons could move back spontaneously and do work on their own merit, then the membrane would not be semipermeable!

[9] If we had only a watermill at hand, then there would be no other options! But what if the watermill was not connected to a pump or generator or any instrument that could do useful work? Do we still indulge in the vanity of pumping the water up, just because we can?

[1010] In the holistic perspective, if such a chemiosmosis process were to be operative, cells would not have energy accountability! The important secondary aftermath of this analysis is that fundamental laws of physics will not permit us to have the same membrane portal working energetically efficient in both directions! Therefore, ATP synthase can only be an ATPase (the clearly proven in vitro function), not a synthase! We know that while some man-made DC motors may function reversibly as DC generators (by electricity-magneticism facile interchange), the chemistry and mechanics of the molecular motors would be quite different.

***b.** Elementary phenomenology recaptured and reassessed:* Let us now consider the elementary aspects of the reaction, as per the prevailing ideas, taking NADH as the simple starting material.

Step 1: NADH $\rightarrow$ NAD$^+$ + H$^-$ (where H$^-$ $\leftrightarrow$ $_m$H$^+$ + 2e$^-$) ………….…….. $\Delta$E = -p

This reaction would be the primary directive source, releasing energy and being spontaneous. It shall become a kinetically viable process ONLY if a suitable acceptor of hydride ion (or proton and electrons) is available nearby, to pull the equation to the right. In principle, the ETC could serve precisely that purpose. The process could liberate non-trappable heat energy too! Further, the energy released therein can be recycled only if suitable modalities of trapping mechanisms are available within the system; and this excited or surcharged or altered state of trapping agents must be a significantly stable one. Further, these altered states should have modalities for relaying the necessary information and/or force to the subsequent element(s) in the machinery.

Step 2: (H$^-$ $\leftrightarrow$ $_m$H$^+$ + 2e$^-$) $\rightarrow$ $_{ims}$H$^+$ ‖ 2e$^-$ …………………………….… $\Delta$E = +x

Here, energy must be expended to achieve the spatial separation across the inner phospholipid membrane, which could be more than 4 nm thick! This energy could be derived from Step 1. Now, I wonder if it would take very different amounts of energy to move a positron, hydrogen ion, and caesium ion across the lipid membrane. I have not seen any calculation factoring this effect that the charge is distributed across a mass which consumes significant space and we are moving the charge/mass through ~40 Å of low dielectric medium. (Quite simply, this is in no way a "unidimensional" phenomenon.) The randomized and non-directional thermal energy cannot discount for the requirement for directional movement of mass across significant space. (This aspect is important because while the proton is being thrown out, the chemiosmotic postulate accounts for the electrical component alone. But while the same proton comes in, it seeks to cash in on its mechanical/electrical attributes. But you will see that this concern is not that important because there are even greater disconcerting aspects to follow!)

Step 3: $_{ims}$H$^+$ $\rightarrow$ $_m$H$^+$ …………………………………….…...…….. $\Delta$E = -y

Now, only a part of the energy released in Step 1 can be harnessed for the proton's return into the matrix phase. The inability of the inner membrane to offer any decent connectivity between the two macroscopic phases does not allow an efficient harnessing of the potential of x either. If this favorable energy term was somehow dependent on Step 2, we can speculate that |-y| << |+x|.

Step 4: $_mH^+ + e^- \rightarrow$ H  **Or**  $_{ims}H^+ \rightarrow {}_oH^+$……………………………..……..... $\Delta E = \pm z$

Some reactions that could either aid or disadvantage the overall flow. The sign of z is kept $\pm$ because it remains uncertain as to what way it could contribute to the previous three steps.

Now, by the law of conservation of energy, we can state that-

$|-y| \approx |-p + x \pm z|$

If we concede that Step 2 (the matrix to intermembrane space proton pumping associated with the electron transfer and electron-proton separation) efficiently harnesses most of the energy released in Step 1 and energy terms of Step 4 are minimal, then $|\pm z| \rightarrow 0$ and $|+x| \rightarrow |$ -p|. As a consequence,

$|-y| \rightarrow 0$

This simple deduction further ratifies our understanding in the discussion of Step 3 that $|-y| <<$ |+x|. Therefore, it would render such an elaborate process (movement of electrons from NADH through protein and small molecule machineries, pumping out and moving in of protons) energetically inefficient. Therefore, the ETC-Chemiosmosis-Rotary synthesis machine cannot be tenable or viable!

*If you bought into the "energetic" mathematical logic demonstration I made above, I must say that I took you for a ride. Here is why- Step 1 and Step 3 cannot be energetically connected through Step 2, violating the assumptions we started off with! (That can also be stated as- Step 3 cannot have a negative sign when Step 2 has a positive one! You may have whatever vectorial arrangement of the module within the inner membrane, but the energy terms would always disagree with one another. Else, the vectorial arrangement would have to change in time, which would incur energy expense. So, we will have to find that energy from elsewhere, messing with the overall accountability, and feasibility of the process to serve as a working machine.) But this is precisely what the prevailing explanations achieve, and I have just used the very same ideas,*

*only to show the vain purpose it leads us to! Yet, the point that I proved in this exercise is that even if we tag along the prevailing explanations, we get nowhere in terms of explaining the energy requirement for a viable machine. We either land up at a super-efficient machine or we cannot explain the energy requirements.*

Now, let me ask you a question after I tell you a small story- Three friends go to a bar and have a drink each. The bill runs to 30 bucks and each one of them shares 10 bucks and thus, 30 bucks is pooled. The waiter takes it to the manager's counter, where the manager gives them a 5 buck discount and puts a fiver back on the check platter, to be handed back to the customers. The waiter knew that the three customers had gone dutch and therefore, he knew they would have difficulty sharing the fiver amongst them. So, he took the fiver and placed three single buck coins on the platter instead, to return to the customers. They duly took one buck apiece. Now, each one of them spent 9 rupees and since there are three of them, the total comes to 27. The waiter has taken 2 bucks. Where did 1 buck go, out of the total of 30?

Did you see how a poorly framed (illogical) question can make a mess of simple accounting principles and take one on a wild goose chase?

*c. The magical mitochondrial membrane that Mitchell wove into existence*: Now, let us get back to proton pumps, or Step 3 above, the most obscure but yet tangible part of Mitchell's proposal. (It is easier to comprehend, than the electrochemical potential being used by ATPase part!) Most known proton pumps expend ATP (because working against a gradient requires some serious energy!). The ETC membrane proteins of mitochondria are supposed to pump protons (three trans-membrane and one into the membrane; all of which supposedly work against gradients!) by virtue of conformational changes or water relays induced/produced within (conformation gated channels). Some say that these membrane proteins achieve the feat by availing "high-energy electrons" from NADH, succinate, reduced CoQ and Cyt. *c* for the respective four complexes. Electrons can have a relatively higher energy only if they are accelerated in some electromagnetic field or are present at higher temperatures! Otherwise, an electron obtained from NADH or reduced Cyt. *c* or ubi(semi)quinone or superoxide are identical. Therefore, the electrons at say Complex I cannot be of "higher energy", than an electron at say, Complex III!

(This elementary fact is justified by the finding that ATP is synthesized at three "sites", disregarding the so-called "energy status" of the electron!) It is incomprehensible how or why the very same electron would release "definite packets of energy" while being transferred to (or across) the Complexes, simultaneously accounting for (the non-existent!) proton pumping feats![11]

The worst part of the whole ETC-proton pump coupling concept is that there exists no direct evidence for this assumption. Acquiring ATP synthesis with a lower pH outside (relative to the matrix pH) is no direct evidence that matrix pumps out protons! It just means that when excess protons are presented to mitochondria, the equilibrium position is shifted towards synthesis. Further, the indirect evidence given for proton pump (increase in pH of a mitochondrial suspension when oxygen is given) must be faulty protocol or misplaced inference! Why use valinomycin-K in this experiment?[12] (Such erroneous protocols will be discussed in a later part.) To me, this could also negate the most fundamental of Mitchell's postulates which requires an intact inner impermeable mitochondrial membrane! Even otherwise, one wonders why the pH elevates in a few seconds and drops quickly back to the original within the first minute itself! Also, showing that the matrix becomes alkaline (by using specific pH indicators) might just imply a production of hydroxide ions in the inside, owing to a reaction (and not necessarily the pumping out of protons, which we know cannot happen!). Let's concede that the oxidation of NADH, succinate or reduced CoQ can release protons and energy, and protons thus generated can potentially be given out into the inter-membrane space or into the phospholipid membrane (for CoQ to recycle). But the terminal stage of Complex IV has no such energy-harnessing machineries or mechanistic options! How can it pump the same number of protons from a deficient matrix side to inter-membrane space (with the same fervor as that of Complex I, where the electrons started their presumed journey) and put up four protons for water formation?.

---

[11] This statement is very solid on scientific logic because the redox potentials of the redox active intermediates do not follow the "beautiful" logic of fixed redox windows, as detailed in the textbooks.

[12] This will be discussed in a later part below. When I see the protocols in the reconstituted systems (in the pertinent research area), I can immediately infer that much of it is clearly misplaced!

But once again, let us clearly see that the chemiosmosis supposition keeps asking for too many mutually contradictory requisites. We want a low proton density at the matrix side to serve the proton gradient and we want a high proton concentration at the same locus to serve the proton pump. It seems like we want to save the cake and have it too! See the farce with Complex IV- four protons are taken from the intermembrane matrix-ward and 4 protons are taken from the matrix outward- all for one oxygen molecule's reduction. What is the molecular logic or energetic drive for such predestined agenda? (Other than meeting the mandate imposed by Mitchell's postulates of chemiosmosis?) In both the automobile engines and hydroelectric power-plants, the existence of an "externally open sink" is a must (the fume exhaust or after-bay outflow, respectively) for ensuring the spontaneity and thermodynamic feasibility for the system to do work. The highly constrained inter-membrane space cannot be deemed as an open outlet (for the expulsion of protons by Complexes I, III & IV). It must not be forgotten that the mitochondria are relatively closed systems. Unless freely soluble and gradient-favored, entry and exit of most species at all points are supposedly taxed with the currency of energy (ATP). How can the proton pumping out machinery be viable in this system? Protons have been pumped out of one macroscopic phase (matrix) to another (inter-membrane space) and they are disconnected by the semipermeable membrane. Why should the protons feel the pressure to go back in now? Perhaps, they have become over-energetic and nostalgic too (of the nice times they had in the matrix!).

One could imagine a protein undergoing structural or conformational changes upon reduction. But it is difficult to see tethered and embedded multi-component protein assemblies repeatedly flipping themselves out or conducting protons through a membrane (against the gradient, when the proton concentration within is in single digits!), without the loss of serious amounts of energy! (All the while, we have to also accept the assumption that the membrane is impermeable, or should I say, "conveniently permeable" to protons!! If Mitchell had used this term, his hypothesis would have had more appeal!) But let us keep indulging this consideration even further. Let us accept that reduction of a protein could reorganize its electronic distribution and therefore, potentially effect a change in its conformation or induce some specific channeling/relays and thereby, enable flipping of protons across membranes. But that would require an undisturbed "surrounding stature" to rest on and considerable time for the process to occur. That is- once a protein gets an electron, there should be no significant amounts of redox active agent(s) nearby, that could mess with (or quench) its 'excited state'. Then, the protein should flip and upon this event, it should lose the proton and thereby revert to its natural state. This requirement too is rather unlikely to be met because the phospholipid interface would be subjected to a barrage of different types of species.[13] Furthermore, the passage of an electron through a given point within a relay could potentially lead to an expulsion of a single proton to the outside. I just cannot fathom how the movement of a single electron could pump 2 protons out, as is advocated for Complexes I & III. Further, some workers also believe that even CoQ could pump protons out! What could such suggestions be deemed as, except "intellectual pandering" to a pre-existing superior peers' demand? Citing such works is not worth my time.

Let me give another analogy here- Four elephants (duly marked with identity tags) from Kulappully (my village in Kerala, the state of elephants in India) keep going to Lake Manasarovar (a dream landscape in Tibet). A Maruti (a popular brand of sedan in India) with Kerala registration plates is also found in Manasarovar, at times. Now, just because the car has four seats and both (car and elephants) originally belonged to Kerala, one cannot have a validated hypothesis (given a theory status!) that the elephants drove down to Manasarovar in the Maruti. Particularly, when we know that the elephants were found in Manasarovar within a couple of hours of their disappearance from Kulappully and the car landed up in Manasarovar only after a couple of days! Even if we don't have a measure of the mass and size of the elephants versus the car, the time discrepancy and the distance involved (>3000 km) are factors which should tell us that the hypothesis is misplaced! The 4 elephants could not have walked or driven there on their own jolly will (not without google-maps to direct them, in that good time!), and if we knew a bit more about the terrain, we would also know that the car cannot be driven all the way there either! Therefore, we must ask the elephants how they went there or find it out ourselves, rather than stick to the Maruti explanation (just because it happens to be the only lame reason we have!).

---

[13] Phosphorylation of membrane proteins achieves conformation and functional changes. These kinds of two-electron reactions (coupled with group transfer) are relatively well established. Their existence is not being challenged here because the time frames and competing species therein are within acceptable levels, to afford feasibility.

While the discussion is on time, let me reiterate that the hardest of all facts for the ETC-proton pump hypothesis to come to terms is with the reality that if ET rates are high in the microsecond ranges, it does not give any conceivable scope for a proton transport to be coupled with this event! Proton transfer across the membrane would take time in the millisecond ranges. But if electron transfer is somehow slowed by biological tuning (to accommodate proton translocation), we cannot explain electron transfer rate! Clearly, the two don't seem to go together! So, both spatial (the site of reduction is several angstroms away from the transmembrane helices supposed to serve as proton pumps!) and temporal (the ET rates are too fast for it to be coupled with transmembrane proton translocation) considerations speak against transmembrane proton pumping with the ETC. (And let me reiterate that I have already argued that the ETC cannot account for the rates even as is- it is too slow with the given circuitous setup!) The ETC was in place to serve a proton pump that never was. Therefore, the ETC was never in place either!

The following discussion is the author's enhanced presentation of the critical insights offered by Ling [1981] and the references sited therein, particularly within the section II D & II E, titled- "Is the mitochondrial inner membrane impermeable?" and "Functions of uncoupling agents and ionophores". If a proton on the inside is not moved across the membrane and if by some "domino effect", a proton-binding event at the matrix end can be instantaneously (within microsecond time ranges) relayed to the inter-membrane side, then the trans-membrane region should have a high dielectric, conducting a relay of protons. In that case, the conductance of this membrane and its permeability to protons would be high, and this would break the fundamental postulate of Mitchell's hypothesis. The inner membrane's reported proton permeability of ~$10^2$ nm/s is not unusually low (as originally claimed by Mitchell!), but is quite similar to several other phospholipid membranes' permeability for protons and other cations. Further, Mitchell's postulate demanded a conservative trans-membrane potential of ~200 mV and relatively high resistance of the membrane ($10^6$ to $10^8$ $\Omega$ cm$^2$). We can envisage that the greater the amount of phospholipid content within a plasma membrane, greater is its resistance/impedance. This rational supposition is duly supported by experimental results. Cells with >70% phospholipids (for example- nerve myelin sheath) within their membranes are known to give surface impedance of ~$10^5$ $\Omega$ cm$^2$, whereas most other normal cells give much lower values, ranging two to six orders of magnitude lesser. By weight, the outer mitochondrial membrane comprises of

~50% phospholipids whereas the value for inner membrane is <20% (high density of embedded proteins constitute the major remaining components). Therefore, we can only expect the status of affairs to counter Mitchell's requirements! The mitochondrial membrane surface impedance was found to be <$10^1$ $\Omega$ cm$^2$, approximately six to eight orders short of the value projected by Mitchell. It is only natural that high amounts of redox-active metalloproteins in the inner mitochondrial membrane (and aquaporins) could only enhance its conductance. The resting potentials across the mitochondrial membrane was more than an order smaller (<20 mV), and at times, even of the opposite polarity (compared to what was dictated by Mitchell's postulate). These experimental findings go well with variations known for other cellular/plasma membranes. Further, using monactin and valinomycin (in conjunction with added potassium ions), it was shown that the diffusion barriers offered by the inner mitochondrial membrane (as perceived) is not owing to a continuum of phospholipid layer. It is known that mitochondrial membrane readily exchanges cations like sodium, potassium or calcium. Further, the inner membrane also has significant and no unusual permeability features with respect to anions; and also houses lots of aquaporin [Calamita et al, 2005]. Therefore, Mitchell's claim that the inner mitochondrial membrane is a lot different from others is completely unfounded. Equally, his methodology to use valinomycin-K as a tool to measure potential across the membrane is surely erroneous because of several unexplained fundamental observations. (It should be noted that the valinomycin induced exchange of internal protons with external potassium ions remains more of an unexplained/neglected phenomenon, even now. For example- Calcium uptake by mitochondria increased with metabolic inhibitors of Complex I, III & V; and also with valinomycin. But addition of high potassium ions lowered accumulated internal calcium! Further, it was found that valinomycin did not just facilitate K$^+$/H$^+$ ion replacement & diffusion-equilibration across the membrane, but the former's concentration determined the equilibrium position of the K$^+$ ion in/out distribution! These findings do not permit the techniques that were/are employed in the area.) The reader may kindly peruse Gilbert Ling's works in this regard to note some interesting experimental observations.

We shall now deal with the summated ideas of the exercise we indulged (last paragraph of point **a** above) of comparing mitochondria with watermill. Regardless of the spatial arrangements into respirasomes or any other type of distribution or connection on the inner membrane, the overall system incorporating the two fundamental modules g (proton pump) and h (ATP synthase) must work through at least three distinct time points (t0, t1 & t2; t3 can be considered equivalent to t0), as shown in Figure 2. For the time being, let us disregard that proton pumping machineries are not a viable option in mitochondria and let us go along with what the erstwhile ideas solicit. This exercise is done so that the reader realizes how the Moebius loop was made, how the reality was broken.

**Figure 2**: Schematic diagram chronicling the sequence of events proposed for a minimal model of the inner mitochondrial membrane

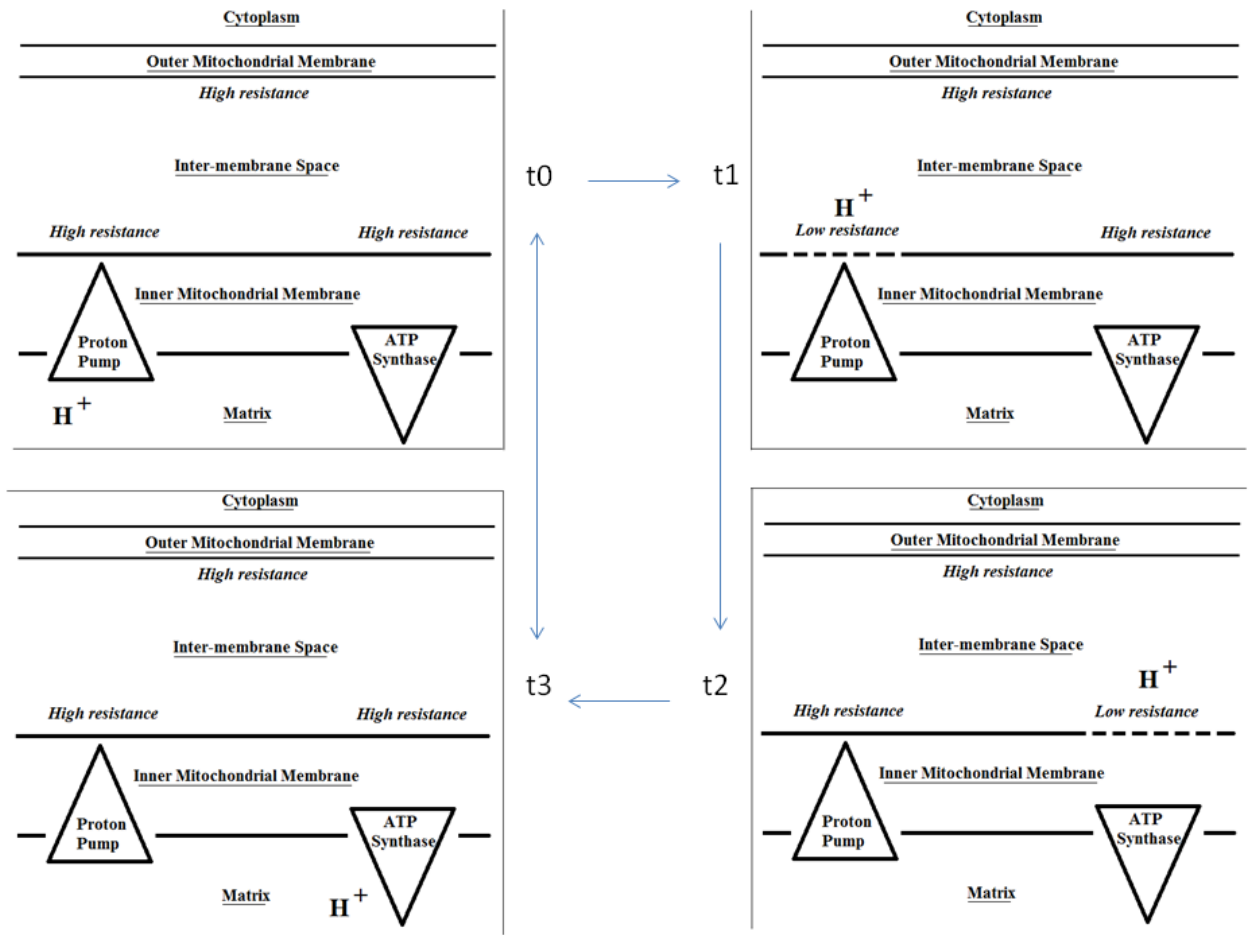

Let us consider the spatio-temporal aspects that govern the phenomenology in these two systems and try to picture the system, starting from simple idealized conditions, imposing constraints that enable us to simplify the system. The simple rendition above (with the assumption that the outer mitochondrial membrane is impermeable to protons and protons are not consumed or lost in the overall process) could help the reader see what the hypothesis achieved in real space/time. (In this image/at this stage, we shall not consider the energy requirements at all.) You will see how the same locus was used both as a permeable and non-permeable point(s). The reader should make a thought exercise and see if opening up the outer mitochondrial membrane could enable a feasible and continuous model. In that case too, we will just end up with a double Moebius! (That is- if after the stage t3, proton gets consumed, then at t4 (there would be no proton in the matrix). In that case, at say t5, the outer membrane must become permeable to enable excess protons to enter the inter-membrane space and drive ATP synthesis. (By introducing protons from outside, we are disregarding the energetic catch that we are making the machine non-viable and unaccountable.) You will see that there is no way the setup can work in a "proton-pump + proton-inlet" mode. This setup is not like a hybrid car that uses an engine to burn hydrocarbons (a source of power) and battery (yet another source of power) to reduce hydrocarbon consumption. From the energetics perspective, it would be like having two engines; protons sourced from the mitochondrion (ETC) and protons sourced from the cytoplasm. Then, what would keep the machine going (if we allow Mitchell's postulate to hold good), are the protons that come from cytoplasm! In this case, we are still left with the catch 22/zugzwang that the outer mitochondrial membrane would need to secure the permeable-impermeable feature, energy equation is still unaccounted and also, the ETC-proton pumping would be rendered a redundant exercise anyway!

Now, it must be evident that the chemiosmosis explanation that linked the ETC-Proton pump and Rotary synthesis modules was an unreal expression. (It is very important for the reader to realize how he/she was taken for a ride all the while!) It served only to justify queries at a given space and time, but not to make spatio-temporal or energetic sense *in toto*, in a holistic way. (Just as the water always seems to flow downhill in Escher's waterfall, just as the ant sees only one plane in the Moebius loop!) A scientific theory must have provable and disprovable facets and Mitchell's doesn't! The fabric of reality was broken and resealed to give the impression of a continuum that never was. Mitchell's concept of the inner mitochondrial membrane served as a Moebius plane that would give favourable energy transduction when being impermeable at the start and permeable in the middle and impermeable again at the end, as the pump/synthase worked (either in a singular scenario or in a hyper-concerted/synchronized scenario).

*d. Testing the quantitative logic:* Now, I MUST resort to overkill. I would ask you to disregard every single thing discussed earlier and let us go purely by the prevailing concepts for the following discussion, which discusses the energetics with numbers/concepts available (as per the prevailing understanding) from the real world.

**I.** If we account for the mechano-energetics of proton pumps alone, we get the maximum efficiency of ~ 53 % for ATP formation. {If we take ~35 kJ/mol as a conservative value for 1 ATP molecule formation from 3 to 4 protons moving in (each with a maximum free energy value of ~22 kJ/mol), we have the efficiency = [35 / (3 **or** 4 x 22)] = 53 to 40 %.}

This process is a subset of the overall redox-coupled chemico-energetics equation, for which we get an overall (average) efficiency of ~48 %. {For the complete oxidation of one molecule each of NADH or $FADH_2$, we get a maximum of 3 or 2 ATP molecules, respectively. Then efficiency = [(3 x 35 / 220) = 48 % and (2 x 35 / 150) = 47 % respectively for NADH and $FADH_2$].}

Therefore, the proton motive force concept affords very little window for any energy loss (even with the best option of 3 protons per ATP) or does not add up to give the "higher efficiency of 53%" (with the lesser option of 4 protons per ATP) observed in the system.

**II.** Further, if we consider that ~33 molecules of ATP are formed per glucose molecule oxidized {Since the number of ATP formed per two electrons going through the ETC is "demonstrated to be maximally between 2.5 to 3 & 1.5 to 2 for the introduction of the electrons at Complexes I & II respectively", we could have anywhere between 30 to 36 ATP molecules formed. Therefore, the value of 33 is taken as a midpoint average.}, the overall efficiency of biological glucose oxidation would approximate to a value of ~39 %. {The complete oxidation of glucose to six molecules each of carbon dioxide and water is associated with a theoretical energy term of 2937 kJ/mol. Then, the overall efficiency is given by (33 x 35 / 2937).} This value approaches the lower end of the energetic yield of the proton pump scheme alone, giving no scope for energy wastage in any other processes involved in glucose oxidation! (The reactions of Glycolysis and Krebs' cycle and all the molecular transports involved have little rooms for their conduction!

And this is when energy losses incurred in the proton pumping out scheme have not been really accounted for!) Quite simply, we have a "non-working machine" in the proton-pump based ATP synthesis idea here and only a significantly higher efficiency process for mOxPhos can explain the overall energetics!

**III.** Though redundant, let's engage in this exercise. As per Mitchell, it is the overall electrical potential that drives the ATPsynthase. Let's check out if the potential generated can give the power to turn the ATPase, based on the information that we have available. Defining electric power as a product of voltage (0.2 V; sought by Mitchell) and current (1.6 x $10^{-13}$ A; derived by $\{10^2 H^+ \times 10^4 s^{-1}\} / 6.24 \times 10^{18}$), we have a power generated that is equivalent to 3.2 x $10^{-14}$ Js$^{-1}$. Now, the stepping torque (I guess for 120°) for ATPase was determined to be ~38 pN.nm [Tanigawara et al, 2012] and we know that this must be generated ~$10^3$ s$^{-1}$ (because ~10 protons give one rotation; so $10^4/10$ gives the number rotations per second). This gives a power requirement of 3.8 x $10^{-17}$ Js$^{-1}$ for the rotation of ATPase. (So, was I wrong with the gut feeling from my college days that a few protons trickling in could not synthesize ATP?) The calculation shows that ~0.1% of the power that was generated is spent for moving the ATPase stalk. The work in the EPCR system is only done by the proton moving across the macroscopic phases and that is used for moving the stalk. Now, where goes the rest of the energy involved/generated? Per Boyer, it goes to push the bound ATP out (which must be incorporated into the energy to move the stalk!) and to enhance protein's affinity for ADP + Pi combine. (I find the last supposition nothing but wishful thinking!) Please see- only a fraction of the energy derived from NADH oxidation can be used to pump protons out. If somehow, this energy is retained by the protons (via a nostalgia that they were associated with a high energy process!?) and channelized into ATP synthesis, only a miniscule amount is utilized. Now, the efficiency calculations don't match up either! How/why is the ATP energetics so unaccountable? [[I don't know if this torque calculation above holds because everything depends on the scenario that would be prevailing in the cellular system, in comparison to the experimental conditions. Attaching an actin filament would introduce significant drag in the experimental system. But in situ, the c subunits' cylinder (attached to the shaft) would have to work against the friction imposed by the lipids of the membrane. If these aspects are not comparable, the experimenter should not have done this as an evidence to show that ATPase is a cyclic/rotary enzyme. I haven't bothered to read the details of the work that I have cited here because honestly, such sophisticated science does not appeal to me. To me, science is done to understand something, not to put up a show! The experiment was formulated to show that ATPase was a rotary enzyme! The problem is I wonder if it could have shown us that it is not! If the experimental setup tethered the $F_1$ unit and forced the $\gamma$ stalk to move and then the researcher calculated the force involved (for moving the filament/stalk), we are into a very bad way of doing science! This is because such a protocol could make even a non-revolving protein rotate! If this was not the case and the researchers had this point covered, we are good!]]

*It is true that anyone could have used the above-mentioned calculations anytime to bring down the ETC-proton pump-chemiosmosis energetics. But you must see that the logic of the above is essentially incorrect! If you bought into the calculations **I** through **III**, I must say that once again, I took you for a ride and you did not catch the logical essence of this manuscript! Why? Clue: It was just like the accounting poser for "the lost one buck" after three friends went for a drink in the hotel and the waiter assumed two bucks' tips! Please excuse my attempts to con you. Please see that science is not just reading a paper and aligning oneself to whatever is written. One should also use one's analytical faculties, to figure out if the writer is just serving you some illogical argument!*

Honestly, I very much doubt if Mitchell realized the goof-up he was getting into! To my understanding, Mitchell was just trying to solve a fascinating problem by cranking up his brainshaft, in his own way, which was- Make a fantastic hypothesis and see if the world fits into it! But it turned out to be unfortunate. I shall tell you a story- A guy named Mr. Botchall wanted to light up a house that he had never seen. He hatched up a plan that he would pump water up to the roof, and then have the water flow down using gravity, and harvest the energy using a turbine generator and use that to light up the house. He put his plans up to the authorities and some guys bought into it and supported his plans. This inspired Mr. Botchall, who bought some crayons and made altered plans to improve the appeal. And lo and behold! When the house was visited, it was lit up! All, including Mr. Botchall continued to believe that the house was lit up because of his scheme! (He thought the authorities had put his plan into action and the authorities thought that Mr. Botchall had executed it himself.) But Mr. Findsome turned up to report that the house was already lit up because of street lights; and that the house only had walls but no roof, there was no water in the vicinity, and there was neither a pump nor any turbine generator anywhere around!

Mr. Botchall's story above should warn us that scientific theories are formed following rigorous testing and substantiation with proof. I wonder why someone did not cross-check before granting this fallacy a "theory" status. Scientific bodies must also have a modality to arrive at canonizing a hypothesis into theory status.

### Item 6: The enzymology of Complex V

Let us recapitulate some basic concepts in enzymology now. Fisher's lock & key hypothesis requires that enzyme has selectivity for the substrate and thus, the enzyme "works" on the substrate. Koshland's induced fit hypothesis requires that binding of substrate can alter enzyme structure and therefore, the substrate "works" on the enzyme. Both these hypotheses are supposed to lead to a "reversible transition state" complex of bound reactants or products in/on the enzyme, which has a lowered activation energy barrier than the uncatalyzed reaction complex. Thereby, the reaction can proceed more efficiently (quicker, depending upon certain experimental variables) to attain equilibrium status (the direction of the reaction depending on the starting concentrations of the components). Generally, an enzyme facilitates a quicker attainment of reaction equilibrium by having greater affinities for the substrates than for the products, so that an evolutionary mandate gets served. Otherwise, there is little point in an enzyme's existence if it were just going to catalyze a reaction in both to and fro directions (at any given state of mixture), leading the system nowhere! But the scenario can change even if there were a definite preference within the enzyme (higher affinity for the reactants that feature on any one side of the equation). If the products have accumulated a lot more than the few miniscule amounts of the substrate that remains, then the products also start acquiring the stature of becoming substrates, in spite of the low affinity that they might pose for the enzyme. In this scenario, the enzyme can (and has to!) serve in a reversible function. The enzyme can also freely serve reversibly in some other scenarios- (i) if the free energy change is very low (thermodynamic drives are not major determinants) and/or (ii) if the enzyme has similar affinities for the substrates and products.

Let us first see $F_1$ATPase for what it has been factually proven to be. It is an ATP hydrolyzing enzyme when isolated and assayed *in vitro*. This makes a lot of sense because the ΔG for this reaction is significantly negative (-35 kJ/mol), under standard biological conditions. In normal aqueous solutions (with plenty of ATP and water molecules), thermodynamics dictate that ATP will proceed to get fully hydrolyzed and thus attain "equilibrium". [[Now, why is ATP hydrolysis favored in normal physiological conditions? This is because- 1. The bonds of hydration for ADP and Pi are stronger than the phospho-anhydride bonds of ATP. & 2. The resonance stabilization of products is more efficient and the charge densities are lower on the products than on ATP. (This is an interpretation of the Gibb's free energy equation, based on the consequence that we observe!) Therefore, any aqueous solution of ATP would ultimately dissociate completely to ADP with the passage of adequate time. *So, we can see that ATP's bonds are not really that strong, after all!* The actual reason that ATP hydrolysis affords a means of doing chemical work is because of the fact that the *in situ* concentration of ATP is significantly higher to that of ADP (the system favors the attainment of equilibrium by hydrolysis of ATP) and any system that is displaced from its equilibrium state can do work (while it tries to achieve equilibrium)!]] Now, though thermodynamics dictate the direction of spontaneity, the kinetics are quite a different thing and this is where enzymes feature. Like all enzymes, ATPase merely hastens this attainment of this hydrolytic equilibrium. How does ATPase do this? If textbooks are to be believed, ATPase's affinity for ATP, ADP and Pi are in the approximate relative ratio ranges of $10^{10} : 10^3 : 10^0$.  At a given instant, let us assume that there are mM - μM concentrations of ATP, ADP and Pi molecules/ions around. Since water is in copious amounts everywhere, we shall not bother about it! (Sidenote: The fact that there is lots of water around is yet another reason why the reaction proceeds in hydrolysis direction!) If we assume second order diffusion limited binding for all substrates and products, we can see that at a given instant, ATP has a greater probability of being found/bound on the enzyme. This means that the attainment of transition state becomes more probable with ATP binding. (Acquirement of transition state can be via Fisher's or Koshland's or both mechanisms; but the ester bond can only be broken/formed by some change sponsored by ATPase.) Now, the transition state (which can be summated as "Enz + ATP ↔ Enz + ADP + Pi") can either let go ATP or ADP + Pi. Since it has high affinity for ATP, ADP + Pi are usually let off! The process can also be seen in another way. The transition state acquirement becomes difficult starting with ADP & Pi because they bounce off the enzyme more than stay on it (and because there are two substrates and the probability that both are bound at the same instant is very low compared to the single high affinity ATP's binding phenomenon). Therefore, the probability that ATP could be formed by the enzyme is low. As an outcome of all considerations, ATPase will only hydrolyze ATP even though equal amounts of ATP, ADP and Pi are present. Thus, though it is via a theoretically reversible mechanism, ATPase works practically in a unidirectional way; and this is owing to the enzyme's preferential disposition.

Now, let me bring a very practical experience to your attention, a classical one that Hawking pointed out- An event can be captured by a motion camera and played backwards. But that does not have any relevance in reality! Though we can see the ceramic fabric of a tea-cup reassemble (from smithereens) when reverse rendered, this cannot happen in reality. For example- imagine that we have a very easily squishy but heavy fruit and when you place the fruit at the intake portal of a simple gravity-aided juicer, the fruit's weight can push a wheel down, dragging the fruit along with it and giving you pulp and juice at the bottom outlet. Now, revert the juicer and try to see if you can get the squishy fruit (at the new bottom) back by adding the pulp and juice (at the new top)! Otherwise, keep the juicer in the same configuration and apply suction at the top, to see if the squishy fruit can be obtained at the top. It just won't work. Infusing greater complexity into the model doesn't achieve the reversibility either- you can get a mechanical fruit-juicer and feed an orange from one end, rotate the wheel clockwise and get the pulp + juice at the other. But you cannot reverse the motion of the wheel and get the pulp (stuck within) to come out of any portal. You have to open the concoction up and clean it out manually! This is because there is a practical limit to reversibility, induced by the unidirectionality of free energy expense and increase in entropy. The case being- ATP hydrolysis is associated with a negative free energy term, something I started out with, in the first place!

Even otherwise, the three binding sites of ATPase cannot be equivalent in strict terms because the α-β dimer of at least one site must interact with b-δ units (on top and the sides) and ε unit at the bottom. Therefore, movement of the shaft is unlikely to give exactly identical effects in all three sites, in both directions. So, imagining ATPase as a perfectly reversible enzyme would be a grave misunderstanding (and purely wishful thinking!), given its own structural attributes. Even if a large amount of protons came in through the $F_o$ portion (in the synthetic mode), it would, at best, lead to futile revolutions of the ATPase (given the binding limitations imposed by ADP + Pi). The most sensible logic for a rotary catalysis works only in the hydrolysis direction. An ATP molecule needs to bind to ATPase, which introduces a movement in the shaft, inducing a movement of 3-4 protons through the $F_o$ module. (Please see that binding of proton at the $F_o$ end does not guarantee ADP + Pi binding at the $F_1$ end for it to work efficiently as a synthase! Besides, the aspartate on the $F_o$ part of ATPase looks too trivial a sensor/hook to latch on to a proton, or for serving as a bidirectional pedal. If it were a rely of amino acids involving at least some crucially located histidine residues, perhaps things could have been better, though not still cutting the acceptable mark, because we know this to be a low proton realm!)

## Item 7: Concluding remarks

The best way to catch a fallacy is to assume that it is true and go on verifying it left, right and centre. Surely, it is bound to blurt out its own falsities. We have seen that affording the best case scenarios for ETC-chemiosmosis-rotary synthesis did not achieve the desired quantitative outcomes that these explanations were supposed to give. Therefore, I have now conclusively argued that the ETC + proton pump is not just disconnected from the useful work step (making the up and down transport of protons a totally futile exercise!), I have also shown that **there is no way that the Steps 1 & 2 OR Steps 2 & 3 would have same signs for the energy terms**! (Please refer point b of Item 5, Supplementary Information.) If it had the same signs, then we would have a super-efficient machine that would violate the laws of physics. (Herein, an analogy from the banking sector would hold good- An entry in a bank ledger that had to be made in the debit side was inserted into the credit side; and that too, of a totally disconnected account!) It has also been shown that the proton pump + ETC would be highly inefficient way of doing things. (The supposition that it would be an event of low mechanistic probability has already been discussed in earlier sections.)

We have now seen that trying to comprehend oxidative phosphorylation with the EPCR hypothesis does not allow us a cohesive/coherent logic or quantitative account, with respect to mechanistic, kinetic or energetic perspectives. This is besides the fact that there are no available answers as to how the system senses electrostatics or concentration differences and how chemico-mechanical signals are relayed or transduced across the relevant protein networks. (The structures of all five major respiratory Complexes are known now. There is nothing in these that could work towards achieving the sensory/regulatory role.) Since the structures were revealed, consensus seemed to have crystallized more based on extrapolation of structure-function correlation. For example- the $F_1$ portion has three copies of the α-β protein and the $F_o$ portion has more than 10 subunits of c protein; and the motor rotates in three strokes of 120 degrees. Therefore, the values that seemed to approach the concepts that support these structural details seemed to be favoured by the researchers! Let's remember that aesthetics has little to with reality or science. If Nikolaus Copernicus had not challenged the prevailing perceptions that sun circled the earth, if Tycho Brahe hadn't made those meticulous astronomical observations and if Johannes Keppler had not perused/believed Copernican theory and Brahe's experimental values, the world (perhaps!) would have been in great darkness now. (Not because sun would refuse to revolve, but because we refuse to resolve, solve and evolve, to see the light of truth and reason!)

But it remains to be seen if the experts would care to look at what I have to say and budge from their stances. Max Planck, the man who quantized an otherwise continuum, was one of the most paradigm-breaking thinkers. He had famously stated that a new idea in science takes roots not primarily because of the force of the same, but more because the opposition to the new idea wanes away in time. (This is a sociological equivalent of the combination of Newton's first and third law.) In spite of this "higher level of awareness", he had unwittingly proven his own statement by opposing Feynman's ideas in QED! I have also realized the Planck's pseudo-wisdom from my own personal experience. My critique on the prevailing CYP mechanisms and floating an alternative explanation has hit the bull's eye with murburn concept. Till date, no one has addressed my works in mXM! This is yet another reason why I have gone the circuitous way to campaign to the uninitiated. I hope that the simple arithmetic and forthright logic/tone used throughout this article serves the research community in their effort to seek better explanations.

## References for Supplementary Information